\documentclass[draft]{agujournal2019}
\usepackage{url} %this package should fix any errors with URLs in refs.
\usepackage{natbib}
\usepackage[inline]{trackchanges} %for better track changes. finalnew option will compile document with changes incorporated.
\usepackage{soul}
\usepackage{microtype}

\usepackage{booktabs}
\draftfalse

\journalname{Machine Learning: Earth}

\begin{document}
%% Suppress "manuscript submitted to ..." running header for preprint
\makeatletter
\renewcommand{\@oddhead}{}
\renewcommand{\@evenhead}{}
\makeatother
%%%%%%%%%%%%%%%%%%%%%%%%%%%%%%%%%%%%%%%%%%%%%%%
%  TITLE
%
% (A title should be specific, informative, and brief. Use
% abbreviations only if they are defined in the abstract. Titles that
% start with general keywords then specific terms are optimized in
% searches)
%
%%%%%%%%%%%%%%%%%%%%%%%%%%%%%%%%%%%%%%%%%%%%%%%

% Example: \title{This is a test title}

\title{From stable online coupling to decade-long climate simulations: A machine learning parameterization for cloud microphysics in ICON}

%%%%%%%%%%%%%%%%%%%%%%%%%%%%%%%%%%%%%%%%%%%%%%%
%
%  AUTHORS AND AFFILIATIONS
%
%%%%%%%%%%%%%%%%%%%%%%%%%%%%%%%%%%%%%%%%%%%%%%%

% Authors are individuals who have significantly contributed to the
% research and preparation of the article. Group authors are allowed, if
% each author in the group is separately identified in an appendix.)

% List authors by first name or initial followed by last name and
% separated by commas. Use \affil{} to number affiliations, and
% \thanks{} for author notes.
% Additional author notes should be indicated with \thanks{} (for
% example, for current addresses).

% Example: \authors{A. B. Author\affil{1}\thanks{Current address, Antartica}, B. C. Author\affil{2,3}, and D. E.
% Author\affil{3,4}\thanks{Also funded by Monsanto.}}

\authors{Ellen Sarauer\affil{1}, Mierk Schwabe\affil{1}, Philipp Weiss\affil{2,3}, Axel Lauer\affil{1}, Philip Stier\affil{2}, Veronika Eyring\affil{1,4}}

\affiliation{1}{Deutsches Zentrum für Luft- und Raumfahrt e.V., Institut für Physik der Atmosphäre, Oberpfaffenhofen, Germany}
\affiliation{2}{Department of Physics, Atmospheric, Oceanic and Planetary Physics,  University of Oxford, Oxford, UK}
\affiliation{3}{European Centre for Medium-Range Weather Forecasts, Bonn, Germany}
\affiliation{4}{Institute of Environmental Physics (IUP), University of Bremen, Bremen, Germany}
% \affiliation{4}{Fourth Affiliation}

%(repeat as many times as is necessary)

% Corresponding author mailing address and e-mail address:

% (include name and email addresses of the corresponding author.  More
% than one corresponding author is allowed in this LaTeX file and for
% publication; but only one corresponding author is allowed in our
% editorial system.)

% Example: \correspondingauthor{First and Last Name}{email@address.edu}

\correspondingauthor{Ellen Sarauer}{ellen.sarauer@dlr.de}

%%%%%%%%%%%%%%%%%%%%%%%%%%%%%%%%%%%%%%%%%%%%%%%
% KEY POINTS
%%%%%%%%%%%%%%%%%%%%%%%%%%%%%%%%%%%%%%%%%%%%%%%
%  List up to three key points (at least one is required)
%  Key Points summarize the main points and conclusions of the article
%  Each must be 140 characters or fewer with no special characters or punctuation and must be complete sentences

% Example:
% \begin{keypoints}
% \item	List up to three key points (at least one is required)
% \item	Key Points summarize the main points and conclusions of the article
% \item	Each must be 140 characters or fewer with no special characters or punctuation and must be complete sentences
% \end{keypoints}

\begin{keypoints}
\item We present stable, decade-long climate simulations of an ML-based cloud microphysics scheme coupled to the Earth system model ICON, trained on global convection-permitting simulations to capture subgrid-scale dynamics-microphysics interactions.
\item With the ML-based scheme, we can eliminate several tuning parameters of the classical graupel parameterization and potentially reduce the scope for compensating errors in future hybrid models; although mean-state biases are not yet reduced.
\item We document a path to decade-long online stability, showing that physical constraints and iterative testing with the coupled model are key and confirm that strong offline performance alone is not sufficient

\end{keypoints}

%%%%%%%%%%%%%%%%%%%%%%%%%%%%%%%%%%%%%%%%%%%%%%%
%
%  ABSTRACT and PLAIN LANGUAGE SUMMARY
%
% A good Abstract will begin with a short description of the problem
% being addressed, briefly describe the new data or analyses, then
% briefly states the main conclusion(s) and how they are supported and
% uncertainties.

% The Plain Language Summary should be written for a broad audience,
% including journalists and the science-interested public, that will not have 
% a background in your field.
%
% A Plain Language Summary is required in GRL, JGR: Planets, JGR: Biogeosciences,
% JGR: Oceans, G-Cubed, Reviews of Geophysics, and JAMES.
% see http://sharingscience.agu.org/creating-plain-language-summary/)
%
%%%%%%%%%%%%%%%%%%%%%%%%%%%%%%%%%%%%%%%%%%%%%%%

%% \begin{abstract} starts the second page

\begin{abstract}
% problem
The representation of cloud microphysics and its nonlinear character and scale-dependence is a remaining source of uncertainty in Earth system models (ESMs). Here, we develop and couple online a machine learning (ML)-based cloud microphysics parameterization with the Icosahedral non-hydrostatic modeling framework (ICON). The primary challenge is achieving numerically stable, long-term online coupling when transitioning from training with km-scale data to application in coarse-scale simulations, where the coupled system encounters atmospheric states and feedbacks not seen during training.
% method
The training data is obtained from a global convection-permitting ICON simulation at 5 kilometer resolution, providing explicitly resolved convective dynamics unavailable at coarse scales. The ML microphysics scheme uses a two-stage design: a classifier to identify active grid cells and a regressor to predict cloud microphysical tendencies. Physical constraints such as enforcing mass positivity and overshoot prevention prove essential for numerical stability in the coupled system.
% findings
We demonstrate that achieving stable online coupling requires enforcing physical constraints and careful dataset curation, and that strong offline performance alone is insufficient. The coupled model maintains numerical stability over decade-long simulations with a performance in reproducing the observed climate comparable to the classical graupel scheme. The ML-based scheme eliminates two microphysics-specific tuning parameters of the classical graupel scheme, directly reducing the scope for error compensation by tuning in future hybrid models, though systematic improvements in long-term mean-state biases are not yet realized.
% conclusion
This study demonstrates that stable, decade-long climate simulations with an ML-based cloud microphysics scheme trained on convection-permitting data are feasible, providing a foundation for future hybrid ESMs.
\end{abstract}

%%%%%%%%%%%%%%%%%%%%%%%%%%%%%%%%%%%%%%%%%%%%%%%%%%%%%%%%%%%%%%%%%%%%%%%%%%%%

\section{Introduction}
% Intro microphysics
Uncertainties in cloud microphysics processes, particularly involving mixed-phase and ice clouds, and the challenge of modeling many nonlinear processes on temporal and spatial scales covering several orders of magnitude remain a major source of error in climate projections \citep{Morrison_2020,Lamb2026}. The cloud microphysics parameterization is a core component of Earth System Models (ESMs), as they describe the formation, growth, and removal of hydrometeors and associated phase transitions. These processes have a strong influence on climate through their impact on cloud lifetime, radiative effects, and the redistribution of water and energy throughout the atmosphere.

% Intro ML on improving ESMs
In ESMs, parameterizations are used to represent the statistical effects of unresolved processes at the grid scale. ML has been successfully used to improve ESMs by replacing or augmenting traditional parameterizations that are based on empirical relationships and physical assumptions \citep{Hafner2025, grundner2025, Heuer2025, Behrens2025, Fuchs_Sherwood_Prasad_Trapeznikov_Gimlett_2024}. ML models are typically trained on short, high-resolution simulations or observational data and then coupled to coarse-resolution climate models, with the aim of reducing long-standing biases in ESMs \citep{Gentine_Eyring_Beucler_2021, Eyring2023a}. Over recent years, learning cloud microphysics parameterizations with ML has become an active area of research \citep{gettelman_2021, Perkins_Brenowitz_Bretherton_Nugent_2024, Seifert_Siewert_2024, Sharma_Greenberg_2025, Ko_Kim_Jo_Jang_Song_Lim_2024}.

% Research Gap
A first line of work has focused on emulating cloud microphysics parameterizations within the same model setup used to generate the training data \citep{gettelman_2021, Perkins_Brenowitz_Bretherton_Nugent_2024}. These approaches primarily aim to improve computational efficiency by accelerating ESMs while closely reproducing the output of existing parameterizations, rather than improving the physical representation of microphysics processes. Beyond a full-scheme emulation, some studies have targeted specific cloud microphysical processes such as ice nucleation \citep{Georgakaki_Nenes_2024, Lamb_2025}, offering a more modular path toward improving process-level realism. Other studies have trained ML models on more detailed microphysics representations using superdroplet box models \citep{Seifert_Siewert_2024, Sharma_Greenberg_2024}. These approaches seek to enhance microphysics realism and reduce biases in coarse-resolution ESMs by using detailed, high-resolution process simulations that are computationally too expensive for application in classical parameterizations for ESMs. However, the transition from superdroplet box models to global climate simulations introduces challenges related to scale separation, boundary conditions, and interactions with dynamics. In addition, the computational characteristics of ML-based superdroplet parameterizations differ substantially from those of classical graupel schemes typically used in ESMs, adding technical complexity when implementing such models in global climate models such as ICON to simulate the highly nonlinear climate system.

% summary offline paper
In our previous work \citep{Sarauer2025}, we developed an ML microphysics scheme trained on convection-permitting ICON simulations (5-km horizontal resolution), then coarse-grained to the 80-km target resolution. This design allowed the ML model to capture nonlinear interactions between subgrid-scale dynamics and microphysics, as explained in \cite{Sarauer2025}. The offline evaluation demonstrated strong predictive skill ($R^2$$>$$0.74$) and physically consistent behavior confirmed through explainable AI analysis. Moreover, this ML-based scheme learns an entire microphysics parameterization including ice processes, which remain a particular challenge in atmospheric modeling \citep{Morrison_2020}, though its behavior and robustness in an Earth system model remained untested.

% research gap online
Previous studies have demonstrated that strong offline performance of ML microphysics schemes does not necessarily guarantee stability once coupled online to a host model, emphasizing the importance of implementing constraints on the ML models \citep{Heuer2025,Behrens2025,Klamt2025, Wang2026}. In aquaplanet experiments, ML microphysics schemes often developed instabilities, whereas random forest models produced bounded predictions and remained stable \citep{harder_2022,Brenowitz_Beucler_Pritchard_Bretherton_2020}. For microphysics, the SuperdropNet emulator of warm-rain collision-coalescence has been successfully coupled to ICON in idealized bubble experiments, demonstrating stable behavior while preserving physical consistency \citep{Arnold2024}. However, despite growing interest, relatively few studies have demonstrated long-term online coupling of ML-based microphysics in realistic climate simulations. \cite{Perkins_Brenowitz_Bretherton_Nugent_2024} demonstrated stable year-long coupling of an ML emulator of an operational microphysics scheme in a global atmospheric model, highlighting that extended online stability is achievable beyond idealized configurations. To date, no study has demonstrated decade-long stable online coupling of a complete ML microphysics scheme including ice processes trained on convection-permitting simulations. Ensuring stability and physical consistency of ML-based microphysics across climate-relevant timescales remains unresolved, particularly for parameterizations trained on convection-permitting simulations \citep{Lamb2026}.

% Overview this work and results
To our knowledge, this study presents the first decade-long stable online coupling of a complete ML cloud microphysics scheme trained on convection-permitting (5$\,$km) data, performing scale transfer to a typical climate model resolution (80$\,$km), addressing a critical and largely unsolved challenge in hybrid climate modeling. This sustained stability is what enables a comprehensive 10-year climate evaluation of such a scheme. We show that achieving online stability requires iterative testing and the incorporation of physical constraints, and that good offline performance alone is insufficient to guarantee robust coupled behavior. The ML microphysics scheme does not yet yield systematic long-term improvements over the classical baseline, though only decade-long coupled simulations make a meaningful climate evaluation possible at all. This provides the community with a rigorously evaluated pipeline that shorter or unstable model runs cannot offer. We evaluate stability, mean-state climate, variability, precipitation characteristics, and computational cost relative to the classical standard ICON microphysics \citep{rutledgehobbs}, and discuss implications for the development of hybrid climate models that combine improved physical realism with computational efficiency \citep{Eyring2023a}. For this, we use a range of ESMValTool diagnostics \citep{Lauer_etal_2026, Eyring_2020} applied to decade-long integrations with the coupled model to assess the performance of the simulated cloud properties and climate by comparison with observations, providing a foundation on which systematic future improvements can be built.

%%%%%%%%%%%%%%%%%%%%%%%%%%%%%%%%%%%%%%%%%%%%%%%%%%%%%%%%%%%%%%%%%%%%%%%%%%%%

\section{Methodology}
% Introductory passage
This study builds on our previous offline study \citep{Sarauer2024,Sarauer2025}, in which we developed an ML microphysics scheme trained on coarse-grained, convection-permitting ICON simulations (version 2.6.7 at 5$\,$km resolution). The ML microphysics scheme is here coupled to the same coarse-resolution ICON configuration (version 2.6.4 at 80$\,$km, i.e. no convection-permitting dynamics) used as the baseline in \cite{Sarauer2025}, with the same set of parameterizations called within ICON, allowing a direct comparison between the offline and online results. The 80$\,$km configuration uses prescribed sea surface temperatures and sea ice concentrations (AMIP-style) and relies on parameterized convection, in contrast to the 5$\,$km training simulations where convection is explicitly resolved. That work demonstrated that an ML microphysics scheme trained on high-resolution dynamics can successfully reproduce microphysics offline, outperforming the coarse baseline model in physical consistency. However, we did not test coupled behavior. Here, we focus on the substantially more challenging task of achieving numerically stable online coupling with ICON over a 10-year integration.

% Difference to Sarauer et al 2025
Achieving stable online coupling required fundamental modifications of the original scheme from \cite{Sarauer2025}: 
\begin{itemize}
\item Enhanced dataset preprocessing including outlier enrichment and stricter filters to handle tail events encountered during long simulations,
\item End-to-end training of the classifier-regressor architecture with physical constraints,
\item Iterative stability testing and tuning of the coupled system.
\end{itemize}
While the overall neural network architecture remains similar to \cite{Sarauer2025}, the preprocessing pipeline, training procedure, and evaluation framework have been substantially revised to address the requirements of stable coupled simulations.

\begin{figure}[htb]
    \hspace{-18mm}
   \includegraphics[width=1.2\textwidth]{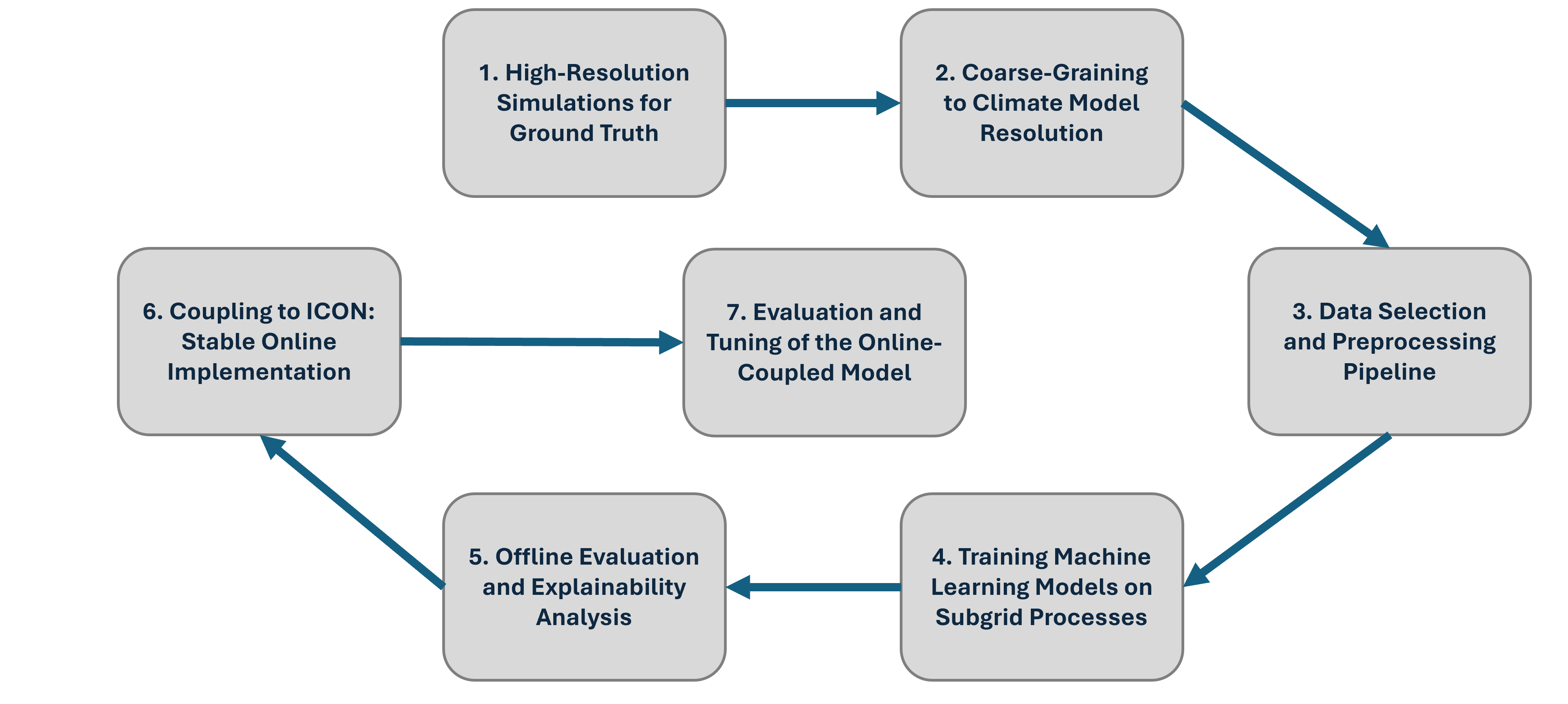} %\vspace{3mm}
    \caption{\label{fig:pipeline}
Schematic overview of the online ML coupling workflow. High-resolution ICON simulations provide training data, which are coarse-grained, preprocessed, and used to train the ML microphysics scheme. After offline evaluation, the ML microphysics scheme is coupled online to ICON, followed by stability assessment, tuning of the online-coupled ICON version, and evaluation of the simulated climate and cloud properties against observations and baseline simulations.}
\end{figure}

% Description of the pipeline
To achieve this, the following methods are designed to prevent problems of ML microphysics schemes when coupled online (e.g., instability and drift due to unphysical results). Figure~\ref{fig:pipeline} summarizes the methodological workflow. The process starts with high-resolution (5 km) ICON simulations that serve as the training dataset. These simulations are coarse-grained to the target climate-model resolution (80 km), after which relevant variables are selected and preprocessed, including data filtering to remove unphysical atmospheric states, outlier enrichment to improve coverage of extreme tendency values (tendency refers to the rate of change (per second) of a variable), and normalization of inputs and outputs by the training set mean and standard deviation. The ML microphysics scheme is then trained and evaluated offline before being coupled online to ICON. Once online coupling is implemented, the coupled system is tuned and evaluated against observations and reanalyses and compared with baseline ICON simulations. This workflow is technically demanding: inconsistencies introduced at early stages may only manifest during long-term online integration, where they can trigger numerical instability. Rigorous validation at each step is therefore essential to ensure physical consistency. For the online coupling, we focus on two primary criteria: numerical stability and accuracy relative to observations. 

\subsection{Dataset}
% Dataset passage (shortened and rephrased from Sarauer2025)
The ML microphysics scheme is trained on data from a global ICON (version 2.6.7 at 5$\,$km resolution) simulation at convection-permitting~resolution. Full details of the simulation setup are provided in \cite{Sarauer2025}. The training data is generated using ICON in the "nextGEMS cycle 3 " setup \citep{Segura_2025}, configured with ~5$\,$km horizontal grid spacing, 90 vertical levels, and a 40$\,$s time step. The simulation spans 22 days (20 January to 9 February 2020), with the first 10 days discarded as spin-up, resulting in 12 days available as training data. Boundary conditions (sea surface temperature and sea ice concentration) are prescribed from the ERA5 reanalysis \citep{Hersbach_2020}, which also provides data for the atmospheric initialization. The high-resolution training simulation employs the single-moment "graupel microphysics" scheme of \cite{rutledgehobbs, cosmo}, which extends the baseline ICON microphysics scheme of \cite{Lohmann1996DesignAP}. While the \cite{Lohmann1996DesignAP} scheme solves prognostic equations for the mass mixing ratios (mmrs) of water vapor, cloud liquid water, cloud ice, and rain, the classical graupel scheme \citep{rutledgehobbs, cosmo} additionally includes snow and graupel as prognostic variables and provides separate precipitation rates for rain, snow, and graupel. Both microphysics schemes also predict temperature changes due to phase transitions (latent heating and cooling due to evaporation).
Table \ref{tab:features} summarizes all input and output used for the training of the ML-based scheme, saved as 3-hour time averages. For the training dataset, model variables are stored immediately before the call to the microphysics scheme, and the corresponding tendencies are stored after calculation of the cloud microphysics and subsequent saturation adjustment. The high-resolution outputs are then spatially coarse-grained to 80$\,$km resolution using the method of \cite{grundneretal}. 
% short comment to limitation of data
Due to computational cost constraints, the training data spans only 12 days in January. A limitation of this relatively short sampling period is that not all possible atmospheric states can be covered. Nevertheless, cloud microphysics processes depend on the local thermodynamic state (temperature, pressure, humidity, hydrometeor content) rather than season of the year or geographical location. The training dataset samples the full global range of these states with temperatures from 180$\,$K to 320$\,$K, ambient pressures from 100$\,$hPa to 1000$\,$hPa, and relative humidities from 0$\%$ to supersaturation. Therefore, we expect good generalization across all seasons and climate regimes, which we verify through decade-long coupled simulations spanning all months (Section \ref{subssec:stability}).
From the 12-day dataset, we select a random subset of 20 million samples (10$\,\%$ of the total dataset) within the first 9 days after spin-up of our simulation for training and validation. We test our method on the 3 remaining days of the simulation. For the final model used in the coupled simulations, we retrain on the full 12-day dataset to maximally exploit the available training data.
\begin{table}
\caption{Overview of all input and output features of the ML microphysics model (mmr: mass mixing ratio). \label{tab:features}}
\begin{tabular*}{\textwidth}{@{\extracolsep{\fill}}lccc@{}}
\hline
\textbf{Type} & \textbf{Short name} & \textbf{Description} & \textbf{Unit} \\
\hline
Input & pf\_mig ($p$) & air pressure & Pa \\
Input & ta\_mig ($T$) & air temperature & K \\
Input & qv\_mig ($m_v$) & water vapor mmr & kg\,kg$^{-1}$ \\
Input & qc\_mig ($m_c$) & cloud liquid water mmr & kg\,kg$^{-1}$ \\
Input & qi\_mig ($m_i$) & cloud ice mmr & kg\,kg$^{-1}$ \\
Input & qr\_mig ($m_r$) & rain mmr & kg\,kg$^{-1}$ \\
Input & qs\_mig ($m_s$) & snow mmr & kg\,kg$^{-1}$ \\
Input & qg\_mig ($m_g$) & graupel mmr & kg\,kg$^{-1}$ \\\midrule
Output & tend\_ta\_mig ($\Delta T$) & tendency of air temperature & K\,s$^{-1}$ \\
Output & tend\_qv\_mig ($\Delta m_v$) & tendency of water vapor mmr & kg\,kg$^{-1}$\,s$^{-1}$ \\
Output & tend\_qc\_mig ($\Delta m_c$) & tendency of cloud liquid water mmr & kg\,kg$^{-1}$\,s$^{-1}$ \\
Output & tend\_qi\_mig ($\Delta m_i$) & tendency of cloud ice mmr & kg\,kg$^{-1}$\,s$^{-1}$ \\
Output & tend\_qr\_mig ($\Delta m_r$) & tendency of rain mmr & kg\,kg$^{-1}$\,s$^{-1}$ \\
Output & tend\_qs\_mig ($\Delta m_s$) & tendency of snow mmr & kg\,kg$^{-1}$\,s$^{-1}$ \\
Output & tend\_qg\_mig ($\Delta m_g$) & tendency of graupel mmr & kg\,kg$^{-1}$\,s$^{-1}$ \\
\hline
\end{tabular*}%
\end{table}

\subsection{Data filtering and physical constraints} \label{sec:constraints}
To ensure physically consistent training samples and numerical stability in subsequent coupling, we apply a sequence of filters and corrections to the training dataset. First, we enforce basic physical validity by removing samples with unphysical atmospheric states such as pressures above 100$\,$hPa, temperatures below 150$\,$K, water vapor mass mixing ratios exceeding 0.2\,kg\,kg$^{-1}$, and negative hydrometeor mass mixing ratios. Next, we apply a mass positivity filter that excludes tendencies leading to negative mass concentrations. At temperatures below the threshold for homogeneous freezing ($T \leq -40\,^\circ\mathrm{C}$), positive tendencies of cloud liquid water and rain are set to zero during preprocessing, preventing the ML model from predicting liquid water formation in the ice-only regime where such processes are physically implausible. Additionally, we remove samples where tendencies would increase the current hydrometeor mass by more than a factor of 10 within a single timestep. In total, this removes $1.15\%$ of the training samples.

\subsection{Outlier sampling and normalization}\label{sec:outlier}
After filtering, we preprocess the dataset to improve representation of extreme tendencies and ensure stable training for the neural network. The target variables for the ML model are the physical tendencies predicted by the classical cloud microphysics parameterization. Their distributions show distinct peaks near zero (\ref{appendix_distribution}), causing standard MSE loss functions to underrepresent extreme values. To improve coverage of the distribution tails, we enrich the dataset with additional outlier samples, consistent with \cite{Yang_Gerber_2026}, who show that tail underrepresentation degrades ML parameterization skill. Specifically, we randomly select 23 million samples from the full dataset of 300 million, and add 2 million outlier-enriched samples identified using a quantile-based approach, selecting samples outside the 10th and 90th percentiles. This strategy provides an improved coverage of extremes and yet numerical stability. We found that excessive outlier weighting biases the model toward predicting unrealistically frequent extreme tendencies, which destabilizes coupled simulations. Additionally, we normalize inputs and outputs by the training set mean and standard deviation to ensure consistent scaling for the MSE-based loss function.
\subsection{Neural network architecture}\label{sec:model}

\begin{figure}[htb]
    \centering
    \includegraphics[width=1\textwidth]{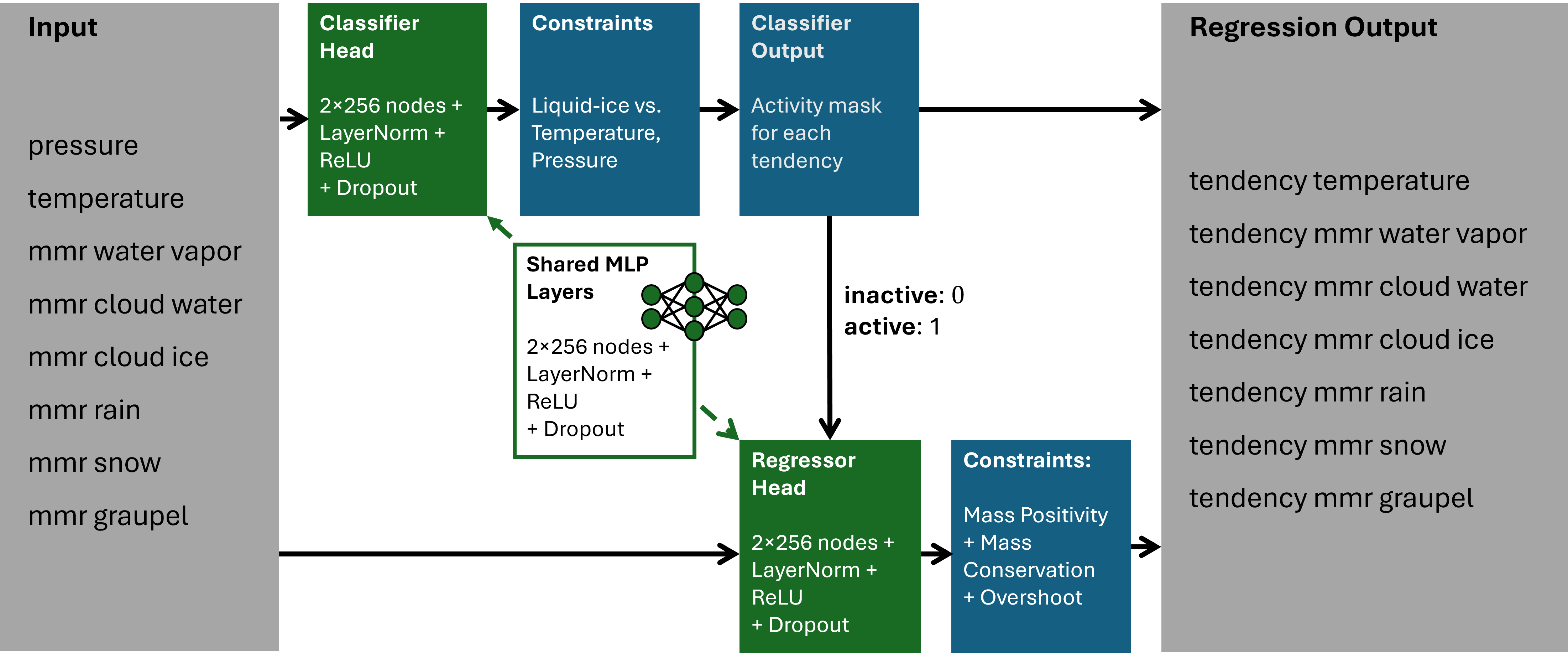} \vspace{3mm}
    \caption{\label{fig:overview_ann} Overview of the ML microphysics scheme. Inputs and outputs correspond to the classical graupel scheme (see Table \ref{tab:features}). The model consists of shared hidden layers, a classifier head to identify active grid cells, a regression head to predict the corresponding tendencies, and post-processing layers that enforce physical constraints.}
\end{figure} 

In contrast to our previous offline study \citep{Sarauer2025}, we develop a combined, multi-head architecture in which a classifier and a regressor are implemented within a single neural network, as illustrated in Figure~\ref{fig:overview_ann}. The model follows a multilayer perceptron design implemented in PyTorch \citep{NEURIPS2019_9015} and is applied independently to each ICON grid cell. The classifier and regressor share a common set of hidden layers, while their task-specific outputs are handled by separate heads. Inputs and outputs are identical to those used in the traditional graupel microphysics scheme to allow for an easy replacement of the classical graupel scheme in ICON with the ML model (see Table~\ref{tab:features}).

% specific architecture details
The shared hidden layers are followed by two task-specific heads. The classifier head predicts whether cloud microphysical activity is expected in a given grid cell and produces an activity mask that gates the regression stage. If the classifier output falls below a prescribed threshold (determined empirically following \cite{Sarauer2025}), all cloud microphysical tendencies are set to zero; otherwise, the regression head predicts the magnitude of the active tendencies. This design is motivated by the fact that in 81$\%$ of grid cells no cloud microphysics takes place, making explicit filtering of inactive cells essential for stable online coupling. A grid cell is classified as active during training when the sum of all hydrometeor tendencies exceeds a threshold of $\epsilon = 10^{-10}\,$kg\,kg${^-1}$\,s${^-1}$),
\begin{equation}
     \sum_q |\Delta m_q| \geq \epsilon ,
\end{equation}
where $\Delta m_q$ denotes the change in the mmr of hydrometeor $q$ per second.  
Both, the classifier and regression heads, use fully connected layers with ReLU activation \citep{relu}, layer normalization, and dropout of 0.1 to reduce overfitting. The classifier head outputs a probability via a sigmoid activation, while the regression head produces linear predictions for all cloud microphysical tendencies. We implement the classifier and regressor as a unified neural network with shared hidden layers and task-specific output heads. This architecture produces an offline skill comparable to training components separately (as tested in \cite{Sarauer2025}) while simplifying implementation. A single model can be loaded and called during coupled simulations rather than managing separate classifier and regressor workflows.

% physics constraints
To guarantee physically reasonable outputs, the regression stage incorporates physics-based constraints. Following \cite{harder_2022}, we enforce non-negativity of hydrometeor masses by constraining the predicted tendencies to satisfy
\begin{equation}
    m_{q,t_{1}} = m_{q,t_{0}} + \Delta t \cdot \Delta m_q \geq 0 ,
    \label{eq:constraint}
\end{equation}
where $m_{q,t_{0}}$ and $m_{q,t_{1}}$ denote the hydrometeor mixing ratios before and after applying the parameterization for a given hydrometeor type $q$ and the model time step $\Delta t$. Additional constraint layers are implemented identically to the preprocessing described in \ref{sec:constraints}.
These deterministic post-processing corrections are essential for ensuring numerical stability during long-term online coupling.

% scaling and hyperparameters
For stable training and coupling, standardized input and output scaling is embedded directly within the PyTorch model. Hyperparameters were selected through a combination of Optuna-based hyperparameter search \citep{akiba2019optuna} and empirical testing. The final configuration employs the AdamW optimizer with a OneCycleLR learning rate schedule, a mean squared error loss, a batch size of 4096, and 50 epochs.

\subsection{Coupling to ICON}
% technical background
The trained neural network was coupled to the ICON model (atmospheric component only, version~2.6.4) using the \texttt{ftorch} interface \citep{Atkinson_2025}. The ML microphysics scheme replaces the full set of equations calculated by the graupel scheme, which is used in the high-resolution model to generate the training data, as well as in the coarse baseline model for the benchmarking. It handles large-scale microphysical processes, while convection is parametrized separately using the Tiedtke scheme in the coarse baseline model. The ML microphysics scheme can be activated via a runtime switch, allowing either the classical or the ML-based scheme to be used without recompiling the code. To ensure reproducibility across the model versions, the implementation was fully containerized. This way, all scaling procedures and physical constraints are applied consistently within the coupled framework. The ML model is applied at each ICON grid-cell, analogously to the classical graupel microphysics scheme. It is called independently for each grid cell in vertical levels below 100$\,$hPa, where microphysical processes can occur. Unphysical values are prevented by the constraints described in Section~\ref{sec:model}. A timer diagnostic was implemented to assess the computational cost of the ML configuration relative to the baseline configuration. Prior to conducting fully coupled simulations, consistency between offline and online inputs and outputs was verified to ensure a correct integration of the ML configuration.

%stability
Ensuring numerical stability of the coupled system was a central goal, as small inconsistencies that are negligible in offline configurations can trigger instabilities during long coupled integrations. To assess whether the coupled system remained numerically stable, we followed a staged procedure. First, short coupled simulations were performed, applying physical consistency checks and monitoring known potential problems such as the accumulation of water vapor. Subsequently, year-long simulations using both the tuned ML and tuned classical configurations were conducted and compared with reference ICON simulations to identify potential inconsistencies. After these tests were completed, a set of ESMValTool \citep{Lauer_etal_2026, Eyring_2020} diagnostics were applied to the ICON output to compare the simulated cloud properties and climate to observations and reanalysis data. For this, 10-year coupled simulations were conducted that were also used to assess the long-term stability of the results.

\subsection{Tuning}\label{sec:methods_tuning}
In order to ensure a fair comparison between the ML microphysics scheme and the classical graupel scheme, the model setups need to be retuned, as any modification of model physics requires retuning to obtain a balanced climate state. Both configurations are tuned using the same methodology and observational targets, with parameters controlling turbulence, convection, cloud cover, and cloud optical properties, though the ML scheme requires fewer tuning parameters as the microphysics-specific parameters of the classical graupel scheme have no equivalent in the learned scheme. 

We follow the history matching approach of \cite{Bonnet_Pastori_2025} to define suitable bounds of the tuning parameters, and subsequently, the automated tuning strategy of \cite{grundner2025}, applying the Nelder-Mead optimization algorithm. 
% history matching
The history matching \citep{Bonnet_Pastori_2025} is applied to the classical graupel configuration to identify physically plausible ranges for the tuning parameters. The procedure uses a Gaussian process emulator trained on short ICON simulations to approximate the model response across the parameter space. Plausibility metrics are computed for each candidate parameter set by comparing the emulated model output against observational targets, and implausible sets are progressively ruled out. This is performed over four iterative waves, each sampling 100 candidate parameter sets from the remaining plausible space using Latin Hypercube sampling, with implausibility assessed at a $65\,\%$ confidence threshold. The resulting plausible parameter ranges (Table~\ref{tab:tuning}) are then used as bounds for the subsequent Nelder-Mead optimization of both model configurations.

% nelder mead
The Nelder-Mead algorithm is a derivative-free simplex optimization method that iteratively adjusts a set of parameter configurations by reflecting, expanding, or contracting the simplex in parameter space to minimize the loss function. For both model versions, we use a tuning optimization window of one month and two starting times, one in January and one in July, in order to account for seasonal variability. To reduce the risk of convergence to local minima or plateaus, five randomly selected initial parameter sets are used for each tuning experiment, and the configuration producing the lowest loss is selected.

% tuning targets
The loss function is constructed from a set of large-scale cloud and radiation metrics, including total cloud ice water path (clivi), total cloud cover (clt), longwave and shortwave cloud radiative effects (lwcre, swcre), outgoing longwave radiation at the top of the atmosphere (rlut), reflected shortwave radiation at the top of the atmosphere (rsut), and net top of atmosphere radiation (rtnt). Each metric is normalized using a min-max scaling based on predefined bounds derived from observations for the months January and July \citep{Bouman2026}, and the total loss is obtained by summing up the individual normalized contributions. This approach ensures that all metrics contribute similarly to the optimization while preserving their physical ranges. The tuning target is derived from multiple observational and reanalysis datasets accounting for some of the observational uncertainty as well as natural (interannual) variability. The tuning loss is defined as the sum of normalized absolute deviations of selected metrics from their observational, monthly reference ranges (upper and lower bounds).

Both ICON model configurations (with ML-based and classical microphysics, respectively) are tuned independently using the Nelder-Mead method. The selected tuning parameters and their corresponding ranges obtained by the history matching method are summarized in Table~\ref{tab:tuning}. They include parameters controlling the parameterizations of turbulence, convection, cloud microphysics, cloud inhomogeneity, and cloud droplet number concentrations over land and sea. The latter are part of the cloud optical properties scheme and are therefore used by both model configurations. The cloud microphysics-related parameters (autoconversion efficiency - ccraut, terminal velocity of cloud ice crystals - cvtfall) are not used in the ML microphysics scheme and are therefore excluded from the tuning procedure of the ML model. The parameter ranges are chosen sufficiently broad to allow meaningful adjustments while remaining within physically plausible limits based on the history matching. The overall goal is therefore not to optimize one scheme relative to the other, but to bring both models as close as possible to the observational data, enabling a fair and objective assessment of whether the ML-based microphysics provides an improvement in performance over the traditional graupel scheme. However, tuning is a non-trivial task, and we cannot exclude that either scheme could perform better under a different tuning strategy.
\begin{table}[htb]
\centering
\caption{Tuning parameters, description and their lower and upper bounds used for the automated tuning procedure.}
\label{tab:tuning}
\begin{tabular}{llll}
\hline
\textbf{Name} & \textbf{Description} & \textbf{Range} & \textbf{Unit}\\
\hline
pr0 & Neutral limit Prandtl number &$0.75, 0.95$&$-$\\
csatsc   & Min. saturation below marine inversion&$0.4,1.05$&$-$\\
crt & Critical relative humidity at top&$0.6,0.9$&$-$\\
crs & Critical relative humidity at surface &$0.86,0.94$&$-$\\
entrmid  & Entrainment rate, mid-level convection&$0.0,2.5 \cdot 10^{-4}$&m$^{-1}$\\
entrpen  & Entrainment rate, penetrative convection &$0.0,2.0 \cdot 10^{-4}$&m$^{-1}$\\
entrdd   & Entrainment rate, cumulus downdrafts &$0.0,2.0 \cdot 10^{-4}$&m$^{-1}$\\
ccraut   & Autoconversion coefficient, cloud droplets to rain &$14.0,26.0$&$-$\\
cvtfall  & Sedimentation velocity coefficient, cloud ice &$1.0,5.0$&m\,s$^{-1}$\\
cinhomi & Ice cloud inhomogeneity factor &$0.4, 1.4$&$-$\\
cinhoml1 & Liquid cloud inhomogeneity factor, stratiform &$0.4, 2.0$&$-$\\
cinhoml2 & Liquid cloud inhomogeneity factor, shallow conv. &$0.4, 1.6$&$-$\\
cinhoml3 & Liquid cloud inhomogeneity factor, convection &$0.4, 1.6$&$-$\\
cn1lnd & CDNC over land, high altitude &$20.0, 80.0$&cm$^{-3}$\\
cn2lnd & CDNC over land, low altitude &$100.0, 250.0$&cm$^{-3}$\\
cn1sea & CDNC over sea, high altitude &$20.0, 90.0$&cm$^{-3}$\\
cn2sea & CDNC over sea, low altitude &$20.0, 150.0$&cm$^{-3}$\\
\hline
\end{tabular}
\end{table}
%%%%%%%%%%%%%%%%%%%%%%%%%%%%%%%%%%%%%%%%%%%%%%%%%%%%%%%%%%%%%%%%%%%%%%
\section{Results}
This section presents the results of the online coupling of ML microphysics experiments with ICON. We first assess numerical stability and identify the configuration needed for robust long-term integration. We then assess how well the ML-based microphysics can be tuned toward observations, compared to the classical graupel scheme. Finally, we assess the physical structure and long-term climate performance using multi-year simulations which are compared to observations.
\subsection{Long-term numerical stability and configuration selection}\label{subssec:stability}
% problem statement
Achieving long-term numerical stability is a fundamental challenge when coupling ML microphysics schemes to climate models. Unlike traditional physics-based schemes, ML approaches require extensive online testing because small changes to training, preprocessing, or inference can cause instabilities that only appear after weeks or even months of simulation. The spatial and temporal structure of these instabilities is illustrated in \ref{appendix_stability}.

% general effect of constraints
We systematically test multiple configurations to identify the necessary conditions for a stable coupling using multi-year simulations. Our experiments indicate three key physical constraints: mass positivity (non-negative mass mixing ratios), mass conservation (total water conservation), and overshoot prevention (limiting tendencies to physically plausible ranges based on training data).
% mass positivity
Mass positivity emerged as the most critical constraint. Without it, simulations crashed within days due to negative moisture values resulting in numerical instabilities. With mass positivity enforced (setting negative predictions to zero), simulations could be run for multiple years. Note that although mass positivity is enforced during training data preprocessing, the neural network can still produce small negative predictions, as it is an imperfect approximation that does not strictly guarantee physical bounds. These are rare but sufficient to trigger numerical instabilities in the coupled system over long integrations. This shows that physical constraints must therefore be enforced both during training and at run-time of the coupled model.
% mass conservation
We also tested strict conservation of the water mass in each grid cell, implemented by redistributing the mass imbalance across all prognostic hydrometeor species (water vapor, cloud water, cloud ice, rain, snow, and graupel), weighted by the inverse standard deviation of each species. While this maintains numerical stability, it can lead to biases in  the vertical cloud structure, as assessed from zonal-mean cross-sections of the cloud fraction showing unrealistic distributions with displaced cloud maxima. This occurs because the classical graupel scheme conserves mass at the column scale through precipitation removal and saturation adjustment, not locally at each grid cell. Given this trade-off, (local) mass conservation was not enforced when producing the results presented below.
% overshoot + general physics 
Overshoot constraints also proved essential for long-term stability. For this, tendencies were constrained to remain within the range observed during training for given atmospheric states. They prevent the ML model from extrapolating into unphysical regimes when encountering rare conditions or distributional shifts. With mass positivity plus overshoot constraints, simulations remained stable for the full 10-year period. These constraints activated infrequently (typically $<$$0.1\,\%$ of grid points for a given timestep), serving as safeguards against tail events rather than fundamentally altering model behavior.

% chosen setup #samples and #train params
We next quantify how training data volume and network capacity affect offline skill and long-term online stability. The upper part of Table \ref{tab:stability} shows that training data volume has a strong effect on both offline performance and online stability. Increasing from 200k to 20M samples improves the offline $R^2$ score from 0.67 to 0.74 while extending stability from 5 days to at least 10 years. This demonstrates a clear relationship between sample size and generalization capability: models trained on insufficient data produce predictions that violate stability constraints in ICON more frequently when encountering the full range of atmospheric states in long coupled simulations. Beyond 20M samples, further increases only lead to marginal improvements in offline skill (0.7461 at 30M vs. 0.7421 at 20M) while maintaining similar stability, suggesting we have reached appropriate data coverage. The lower part of Table \ref{tab:stability} shows the network capacity at a fixed training data volume (20M samples). Undersized networks lack the representational capacity for stable coupling: the smallest network (6.7k parameters) achieves only 2 months of stability despite a reasonable offline $R^2$ score (0.68). Stability increases monotonically with network size up to 340k parameters. The largest network (1.3M parameters) matches the stability of the 340k configuration (both achieve stable 10-year simulations) while showing nearly identical offline skill. This suggests that beyond a certain threshold capacity, additional parameters neither improve representation quality nor degrade stability through overfitting, consistent with the use of dropout regularization and the large training dataset. We select the 20M sample, 340k parameter configuration for all subsequent analyses. This represents the minimum requirement for decade-long stability given the setup in our study: configurations with fewer samples ($<$10M) or insufficient network capacity ($<$87k parameters) fail within a couple of years despite achieving an offline $R^2$$>$0.68. This demonstrates that offline skill alone is not a good measure for online stability.
\begin{table}[htb]
\centering
\caption{Overview of training configurations, offline skill and online stability for the ML-based cloud microphysics parameterization. Stability duration represents the length of coupled simulations without numerical instabilities. The model marked in bold is the chosen configuration.}
\label{tab:stability}
\begin{tabular}{lrrrr}
\hline
\textbf{\# samples} & \textbf{\# trainable parameters} & \textbf{$\mathbf{R}^2$ on test} & \textbf{stability duration}\\
\hline
200k &340 248& 0.6735 & 5 days \\
1M & 340 248& 0.6937& 4 months \\
5M &340 248& 0.7215 & 1.3 years \\
10M &340 248& 0.7328 & 6 years\\
\textbf{20M} &\textbf{340 248}& \textbf{0.7421} & \textbf{$\geq$10 years} \\
30M &340 248& 0.7461 & $\geq$10 years \\
\hline
20M & 6 712 & 0.6812 & 2 months \\
20M & 23 248 & 0.7016 & 9 months \\
20M & 87 504 & 0.7133 & 2 years \\
\textbf{20M} &\textbf{340 248}& \textbf{0.7421} & \textbf{$\geq$10 years} \\
20M & 1 345 560 & 0.7414 & $\geq$10 years \\
\hline
\end{tabular}
\end{table}

\subsection{Tuning}
Both the baseline and ML-microphysics configurations were tuned independently using the identical procedure described in Section~\ref{sec:methods_tuning}. Here, we report the resulting model performance and discuss implications for the two schemes. We conducted the tuning using both January and July conditions to include winter and summer conditions, ensuring that the tuning parameters support stable operation across the whole annual cycle. The resulting configuration (Table~\ref{tab:tuning_results}) remains numerically stable in decade-long simulations. Compared to the baseline configuration, the ML configuration exhibits a higher overall tuning loss, with the difference dominated by July conditions, while performance in January is comparable. The classical graupel scheme, by contrast, exhibits more uniform tunability across seasons, likely because its physics-based formulation is not constrained by a training dataset and can therefore be tuned equally well for both January and July conditions.

\begin{table}[htb]
\centering
\caption{Tuning performance for January and July. Lower values indicate better agreement with observational targets, with 0 indicating perfect agreement. The loss is defined as the sum of normalized absolute deviations from observational bounds for each metric and month (see Section~\ref{sec:methods_tuning}).}

\label{tab:tuning_results}
\begin{tabular}{lrrr}
\hline
\textbf{Model} & \textbf{January loss} & \textbf{July loss} & \textbf{Total loss}\\
\hline
ML microphysics &0.14 & 0.47 & 0.61 \\
Classical graupel scheme & 0.14 & 0.21 & 0.35  \\
\hline
\end{tabular}
\end{table}

These results demonstrate that the ML microphysics scheme's sensitivity to the environmental conditions creates both an opportunity and a limitation for tunability. While the model can achieve excellent agreement with observations when tuned on conditions similar to its training distribution (January), this same sensitivity produces poorer performance when confronting atmospheric states not considered during training (July). The classical graupel scheme, by contrast, exhibits more uniform tunability across the seasons, likely because its physics-based formulation generalizes more robustly beyond the specific conditions used for parameter fitting. This suggests that the ML model has overfit to January-specific atmospheric regimes rather than learning season-invariant cloud microphysical relationships, though this remains a hypothesis that should be tested with a training dataset including July conditions. The tuned parameter values for both schemes are provided in Table~\ref{tab:tuned_params} and are used for all subsequent climate simulations. The best values found for several tuning parameters differ substantially between the two schemes. Most notably, the ML configuration converges to higher critical relative humidity thresholds and cloud droplet number concentrations over land, and lower convective entrainment rates. This reflects systematic differences in how the two schemes interact with the ICON model, and suggests that the tuning partially compensates for a positive cloud cover bias in the ICON configuration with the ML microphysics scheme, particularly in the tropical upper troposphere, rather than correcting the underlying microphysical processes themselves. This interpretation is supported by the geographical distribution of cloud cover biases discussed in Section~\ref{sec:results_maps}.
\begin{table}[htb]
\centering
\caption{Set of best tuning parameter values for the classical (graupel) and ML microphysics schemes sorted by namelist.}
\label{tab:tuned_params}
\begin{tabular}{llll}
\hline
\textbf{Parameter} & \textbf{Classical graupel scheme} & \textbf{ML microphysics scheme} & \textbf{Unit} \\
\hline
pr0      & 0.910 & 0.936 & $-$ \\
entrmid  & $2.10 \cdot 10^{-4}$ & $1.00 \cdot 10^{-4}$ & m$^{-1}$ \\
entrpen  & $3.20 \cdot 10^{-4}$ & $0.70 \cdot 10^{-4}$ & m$^{-1}$ \\
entrdd   & $4.10 \cdot 10^{-4}$ & $4.90 \cdot 10^{-4}$ & m$^{-1}$ \\
cvtfall  & 2.186    & inactive & m\,s$^{-1}$ \\
ccraut   & 16.697   & inactive & $-$ \\
cinhomi  & 0.846  & 0.484 & $-$ \\
cinhoml1 & 0.865  & 1.144 & $-$ \\
cinhoml2 & 0.390  & 1.434 & $-$ \\
cinhoml3 & 0.832  & 1.011 & $-$ \\
cn1lnd   & 21.0   & 55.9  & cm$^{-3}$ \\
cn2lnd   & 177.6  & 246.2 & cm$^{-3}$ \\
cn1sea   & 20.2   & 55.4  & cm$^{-3}$ \\
cn2sea   & 81.9   & 37.2  & cm$^{-3}$ \\
crs      & 0.868  & 0.928 & $-$ \\
crt      & 0.620  & 0.712 & $-$ \\
csatsc   & 0.750  & 1.009 & $-$ \\
\hline
\end{tabular}
\end{table}

\subsection{Evaluation of simulated cloud properties and climate with ESMValTool}\label{sec:evaluation}
To evaluate the performance on climate-relevant timescales of the ML microphysics scheme, we conduct systematic comparisons against observational datasets and reanalysis products using the ESMValTool software package \citep{Eyring_2020}. In the following, we use the reference observational datasets similar to ESMValTool-based ICONEval \citep{schlund_2026_iconeval, Lauer_etal_2026}, see \ref{appendix_obs_datasets}, Table \ref{tab:global_means_refs} for a full list of datasets and references. We assess simulations from the ICON model with both the ML microphysics scheme and the classical graupel scheme over a 10-year period (1979-1988) to determine whether the ML approach produces realistic statistics of relevant cloud and climate parameters suitable for long-term climate projections.
Our evaluation has the goal to examine different aspects of model performance. Models contributing to the Coupled Model Intercomparison phase 6 (CMIP6) \citep{cmpi6} are used to benchmark the ICON simulations and to put the results into the context of other state-of-the-art climate models. Specifically, we use global mean metrics to assess overall accuracy, temporal evolution is evaluated with Hovmöller diagrams, vertical structure through zonal means (latitude vs. pressure), and spatial patterns through global maps. The aim is to identify both the strengths and weaknesses of the ML microphysics scheme relative to the classical graupel scheme and systematic biases against observations.
\subsubsection{Global mean climate metrics and CMIP6 comparison}
\begin{figure*}[htb]
    \centering
    \includegraphics[width=\textwidth]{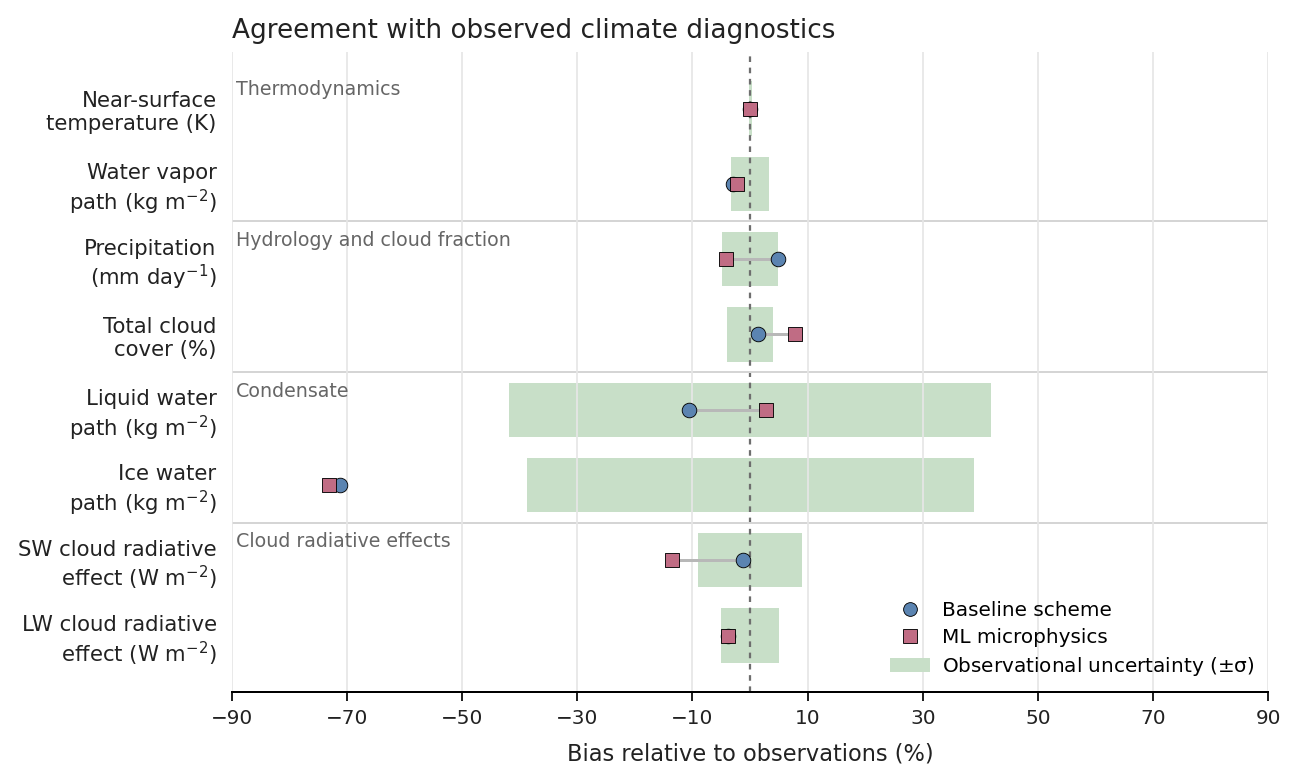} 
    \caption{\label{fig:bias} 10-year global annual mean bias relative to observations for key climate variables, expressed as deviations normalised by the observational spread $\sigma$. Blue circles show the tuned ICON baseline configuration using the classical graupel scheme and red squares show the tuned ML microphysics configuration. Shaded bands denote the observational range (annual mean $\pm\,\sigma$), where the mean and standard deviation $\sigma$ are computed across multiple observational datasets following \cite{Lauer_2022}.}
\end{figure*}

\begin{figure*}[htb]
    \centering
    \includegraphics[width=\textwidth]{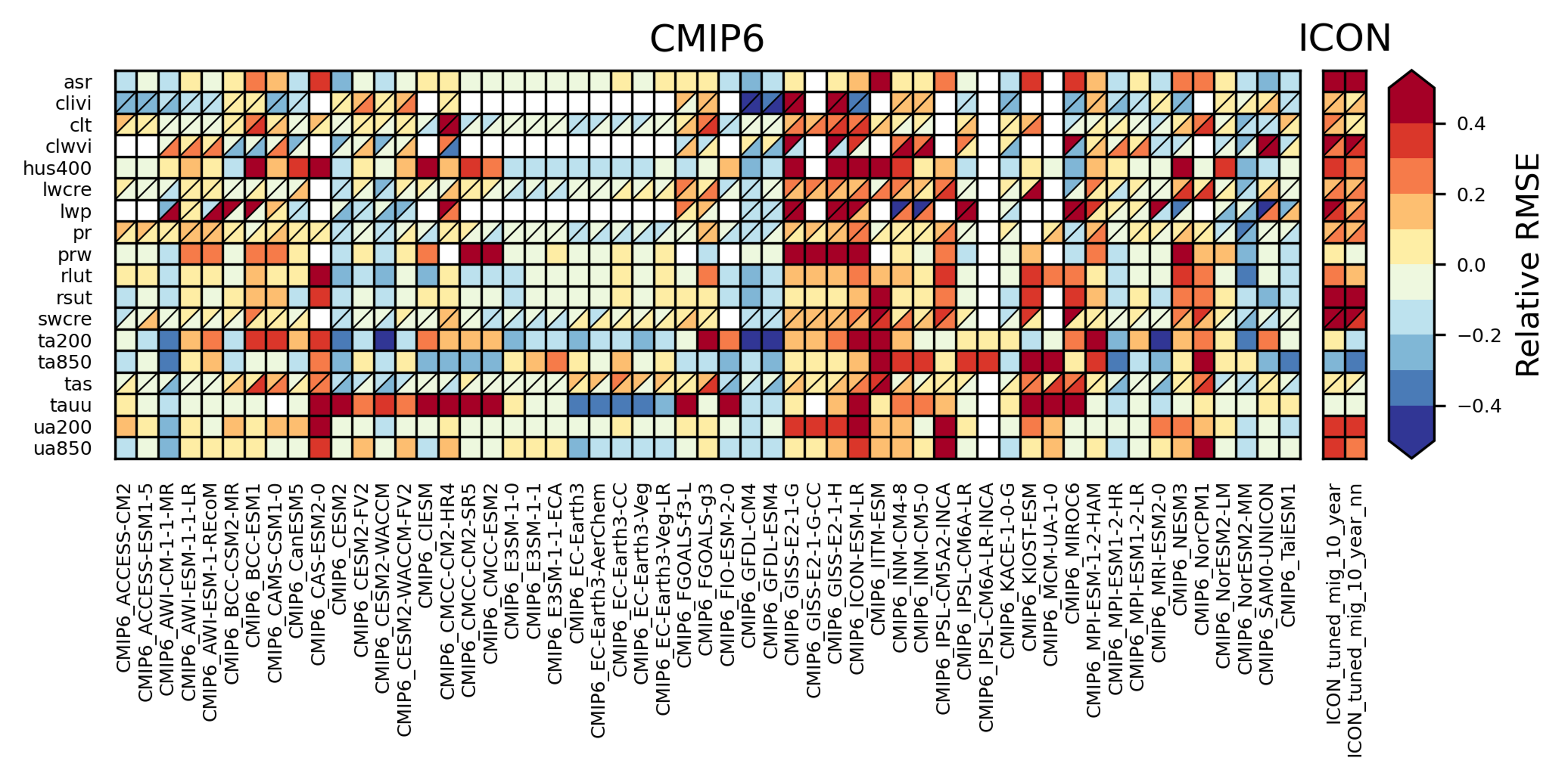}
    \caption{Portrait diagram showing the relative root-mean-square error (RMSE) of selected key atmospheric and surface variables (see Appendix \ref{appendix_names}, Table \ref{tab:cmip6_vars}) across an ensemble of CMIP6 models and the two ICON simulations compared to observational datasets. Rows denote variables and columns individual models. RMSE values are shown relative to the median RMSE from the CMIP6 ensemble with blue (red) colors indicating better (worse) performance than the median RMSE. ICON results from this study are shown separately on the right for comparison.
    }
    \label{fig:portrait_plot}
\end{figure*}

We begin our evaluation by examining global mean values of key variables (Table~\ref{tab:global_means} and Figure~\ref{fig:bias}) and comparing the ICON model performance to the CMIP6 ensemble (Figure~\ref{fig:portrait_plot}). These diagnostics provide a high-level assessment of whether the ML microphysics scheme maintains climate fidelity comparable to established models.
\begin{table}[htb]
\centering
\caption{Global multi-year annual mean values of key climate variables for the baseline configuration, the ML configuration, and of reference datasets.  The observed mean is calculated as the average across multiple observational and reanalysis datasets; see Table~\ref{tab:global_means_refs} in Appendix~\ref{appendix_obs_datasets} for details. The observed range ($\pm 1\sigma$) is shown in parentheses across the reference observational and reanalysis products.}
\label{tab:global_means}
\begin{tabular}{lllll}
\hline
\textbf{Variable} & \textbf{Baseline} & \textbf{ML configuration} & \textbf{Observed mean ($\pm\sigma$)} \\
\hline
Near-surface air temperature (K)                      & 287.3 & 287.2 & $287.2\ (286.4,\ 287.9)$ \\
Water vapor path (kg m$^{-2}$)                        & 24.3  & 24.5  & $25.0\ (24.2,\ 25.9)$ \\
Precipitation (mm day$^{-1}$)                         & 2.8   & 2.6   & $2.7\ (2.6,\ 2.8)$ \\
Total cloud cover (\%)                                & 64.4  & 68.5  & $63.5\ (61.0,\ 66.0)$ \\
Cloud liquid water path ($10^{-2}\,$kg m$^{-2}$) & $6.0$ & $6.8$ & $6.7\ (3.9,\ 9.5)$ \\
Cloud ice water path ($10^{-2}\,$kg m$^{-2}$)   & 1.2   & 1.1   & $4.0\ (2.5,\ 5.6)$ \\
Shortwave cloud radiative effect (W m$^{-2}$)         & -51.3 & -57.6 & $-50.7\ (-55.2,\ -46.1)$ \\
Longwave cloud radiative effect (W m$^{-2}$)          & 25.1  & 25.1  & $26.1\ (24.8,\ 27.4)$ \\
\hline
\end{tabular}
\end{table}
% portrait plot as intro how well models perform

As shown in Table~\ref{tab:global_means} and Figure~\ref{fig:bias}, both ICON configurations produce global mean near-surface temperature, precipitation and water vapor path very close to observations, indicating reasonable global energy balance. Near-surface temperature is reproduced within $+0.12$\,K (baseline) and $+0.02$\,K (ML configuration) of the observed mean, and water vapor path within $-0.73$\,kg\,m$^{-2}$ (baseline) and $-0.55$\,kg\,m$^{-2}$ (ML configuration). Precipitation is underestimated by $-0.07$\,mm\,day$^{-1}$ (baseline) and $-0.11$\,mm\,day$^{-1}$ (ML configuration). For total cloud cover, the ML configuration shows a systematic overestimation of $+4.97\,\%$ relative to the observed mean, while the baseline is closer to observations ($+0.92\,\%$). The classical graupel scheme underestimates global annual mean cloud liquid water path by $-0.70 \times 10^{-2}$\,kg\,m$^{-2}$ compared to the observational mean, while the ML configuration is closer to observed values ($+0.18 \times 10^{-2}$\,kg\,m$^{-2}$). Both schemes substantially underestimate the cloud ice water path (baseline: $-2.86 \times 10^{-2}$\,kg\,m$^{-2}$, ML: $-2.94 \times 10^{-2}$\,kg\,m$^{-2}$ relative to the observational mean of $4.02 \times 10^{-2}$\,kg\,m$^{-2}$), noting that satellite observations of cloud ice and liquid water path are subject to large uncertainties and show large differences across different datasets. The shortwave cloud radiative effect shows corresponding biases, with the ML configuration showing a larger negative bias ($-6.90$\,W\,m$^{-2}$) compared to the baseline ($-0.61$\,W\,m$^{-2}$), consistent with the larger cloud cover. The longwave cloud radiative effect shows smaller differences, with both schemes showing a bias of $-1.00$\,W\,m$^{-2}$ relative to the observed mean of $26.06$\,W\,m$^{-2}$.

The comparison with CMIP6 models (Figure~\ref{fig:portrait_plot}) provides another way of assessing how ICON performs relative to the multi-model ensemble. The portrait plot displays the relative RMSE of the mean seasonal cycle averaged over all grid cells for multiple variables across all CMIP6 models, with red indicating higher errors and blue indicating lower errors compared to the multi-model median RMSE. Both ICON configurations show mixed performance across variables. For temperature-related fields (2-m temperature  (tas), air temperature at 200 hPa (ta200) and 850 hPa (ta850), and zonal wind stress (tauu)), the ICON configurations evaluated here exhibit comparable or slightly better performance than many CMIP6 models (blue to neutral colors). For cloud-related variables (total cloud cover (clt), total cloud water path (clwvi), TOA longwave cloud radiative effect (lwcre)), both ICON schemes show higher RMSE values than the median RMSE (red colors). Notably, the classical baseline and ML configurations show similar RMSE values across most variables, with the ML configuration showing slightly elevated RMSE for some cloud-related fields. Both of our ICON configurations perform comparably to the weaker models in the CMIP6 ensemble even though this comparison is not entirely fair as our ICON simulations use an AMIP configuration with prescribed sea surface temperatures and sea ice, while most CMIP6 models are fully coupled ESMs. However, this comparison provides a rough estimate of what can realistically be expected from a state-of-the-art climate model. Hence, the ML configuration does not systematically improve the model performance over the baseline in this ensemble benchmark.   

Overall, the global mean metrics and CMIP6 comparison demonstrate that the ML microphysics scheme maintains climate fidelity comparable to both the classical baseline and the broader CMIP6 ensemble. While the ML configuration shows degraded performance for some cloud-related metrics, it also does not introduce significant new biases and remains within the typical range of established climate models. This is notable as the tuning parameters available in the classical microphysics scheme are no longer available in the ML-based scheme, reducing the potential of error compensation by tuning in the ML configuration of ICON.

\subsubsection{Hovmöller comparison timeseries }
%Hovmöller plot (5-10) clt, pr, sum of liquid water path ice water path clwvi (variabilität größen)
Figure~\ref{fig:hovmoller} shows Hovmöller diagrams of zonal-mean precipitation and near-surface temperature for the first 4 years of the ICON simulations (1979-1983). Both schemes capture the observed seasonal shift of the Intertropical Convergence Zone (ITCZ) and the annual cycle. For precipitation, the classical graupel scheme shows persistent negative biases in the Northern Hemisphere (NH) Tropics and positive biases in the Southern Hemisphere (SH) Tropics, particularly during NH summer months. The ML microphysics scheme produces qualitatively similar bias patterns but with slightly higher magnitude. For near-surface temperature, both schemes show cold biases at NH and SH mid-to-high latitudes in winter and a slight overestimation of temperature in all other regions, across all seasons.
Neither scheme shows evidence of systematic drift or growing biases over the 5-year period shown, consistent with the decade-long stability demonstrated in Section~\ref{subssec:stability}. The persistent but stable bias patterns suggest that these are structural biases inherent to the basic ICON model configuration rather than signs of numerical instability.

\subsubsection{Zonal means}\label{sec:zonal_means}
% deep dive: zonal means/profiles (hus, cl,) (cli, clw)
We examine the vertical structure of cloud and thermodynamic variables through zonal-mean cross-sections averaged over the 10-year simulation period (Figures~\ref{fig:vertical_cloud_condensate} and~\ref{fig:vertical_humidity_cloudcover}). These diagnostics show how well each ICON version reproduces the observed vertical distribution of cloud-related quantities.

For specific humidity (Figure~\ref{fig:vertical_humidity_cloudcover}, top row), both schemes show nearly identical bias patterns relative to ERA5, with small biases throughout most of the troposphere (baseline: 
$+6.39\times10^{-5}$\,kg\,kg$^{-1}$, ML: $+7.52\times10^{-5}$\,kg\,kg$^{-1}$, 
less than 1\% relative to the ERA5 mean of $1.72\times10^{-3}$\,kg\,kg$^{-1}$). The high spatial correlation ($R^2 = 0.99$ for both schemes) indicates that both configurations capture the large-scale zonal mean vertical humidity distributions accurately. Cloud cover (Figure~\ref{fig:vertical_humidity_cloudcover}, bottom row) shows more substantial differences. The classical graupel scheme exhibits a positive bias of up to 25$\,\%$ in the tropical upper troposphere (region around 100-200 hPa near the Equator). The ML configuration produces a similar but stronger positive bias in this region, reaching about 30$\,\%$ resulting in a higher RMSE and lower correlation. Both schemes show negative biases in the tropical and subtropical lower troposphere. The degraded performance of the ML configuration in reproducing the ERA5 cloud cover patterns suggests that the learned microphysics interact differently with the diagnostic cloud cover in ICON than the classical graupel scheme, despite similar humidity fields. A likely contributing factor is the cell-based architecture of the ML microphysics scheme, which cannot explicitly learn column-integrated conservation properties that the classical graupel scheme enforces implicitly through precipitation removal and saturation adjustment. This may lead to locally elevated cloud water mass mixing ratio tendencies that consequently lead to higher cloud cover than the classical graupel scheme, a pattern that is also reflected in the geographical distribution of total cloud cover discussed in Section~\ref{sec:results_maps}.

For cloud ice mass mixing ratio (Figure~\ref{fig:vertical_cloud_condensate}, top row), observations from CALIPSO-ICECLOUD \citep{NASA/LARC/SD/ASDCCALIPSO} show peak values in the tropical upper troposphere. The classical graupel scheme underestimates cloud ice, showing negative biases in mid-latitude regions. The ML configuration exhibits qualitatively similar bias patterns with slightly higher RMSE and lower correlation values. Both schemes capture the basic vertical structure of ice clouds but underestimate the average cloud ice content. Cloud liquid water (Figure~\ref{fig:vertical_cloud_condensate}, bottom row) from CloudSat \citep{Stephens_etal_2002, Stephens_etal_2018} shows maxima in the mid-latitude storm track regions around 700-900 hPa. The classical graupel scheme shows negative biases throughout most of the troposphere (global mean bias = $-3.58\times10^{-6}$\,kg\,kg$^{-1}$, RMSE = $1.47\times10^{-5}$\,kg\,kg$^{-1}$, $R^2 = -0.25$). The ML configuration shows similar negative biases with comparable RMSE ($1.60\times10^{-5}$\,kg\,kg$^{-1}$) and a similarly poor correlation ($R^2 = -0.49$). The negative $R^2$ scores indicate that both schemes have difficulties to reproduce the observed zonal mean pattern of the observed cloud liquid water distribution.

Overall, the zonal-mean distributions show that both microphysics schemes produce comparable vertical structures with similar bias magnitudes (specific humidity: baseline RMSE = $2.94\times10^{-4}$\,kg\,kg$^{-1}$, ML RMSE = $3.05\times10^{-4}$\,kg\,kg$^{-1}$; cloud cover: baseline RMSE = 6.0\,\%, ML RMSE = 7.1\,\%; cloud ice: baseline RMSE = $1.23\times10^{-6}$\,kg\,kg$^{-1}$, ML RMSE = $1.37\times10^{-6}$\,kg\,kg$^{-1}$) and patterns. The ML configuration does not systematically improve upon the classical baseline in these comparisons, consistent with the tuning results showing limited generalization beyond the training period. The fact that both schemes show similar deviations in cloud cover and liquid water from the reference datasets suggests that these biases arise at least partly also from interactions with other parameterizations rather than from the microphysics formulation alone.

\subsubsection{Maps}\label{sec:results_maps}
% cloud maps bias discussion: reference data, bias1, bias2 (create one big figure) (precip, tas, int. qv(prw)), (lwp, iwp, clt)
We examine the geographical distribution of selected climate-relevant variables through global maps averaged over the whole 10-year simulation period (Figures~\ref{fig:maps_tas_pr} and~\ref{fig:maps_lwp_iwp}). These diagnostics assess how well each microphysics scheme reproduces observed spatial patterns of near-surface temperature, precipitation, water vapor path, and vertically integrated cloud properties.

For near-surface temperature (Figure~\ref{fig:maps_tas_pr}, top row), both schemes show biases of up to $\pm 4$ K with RMSE values of roughly 1-2$\,$K and high pattern correlations ($R^2$ = 0.99). Warm biases appear over the Arctic and some continental interiors (e.g. central Asia, northern Africa, and northern Canada), while cold biases occur over Antarctica and parts of the Southern Ocean. The geographical pattern of the near-surface temperature biases is nearly identical between the two schemes, indicating that surface temperature is determined to a large degree by radiation, dynamics, and surface processes rather than cloud microphysics. The observed precipitation (Figure~\ref{fig:maps_tas_pr}, middle row) shows maxima in the inter-tropical convergence zone (ITCZ), mid-latitude storm track regions, and tropical warm pool regions. The classical graupel scheme and the ML configuration produce biases ranging from $-3$ to $+3$ mm day$^{-1}$ with RMSE values of 1.19\,mm\,day$^{-1}$ (baseline, $R^2 = 0.65$) and 1.28\,mm\,day$^{-1}$ (ML configuration, $R^2 = 0.53$). Notable features include dry biases over the Maritime Continent and parts of the tropical Pacific, and wet biases in some regions such as the tropical Indian Ocean and parts of the Southern Ocean. Both schemes struggle to accurately represent the precipitation distribution in the Tropics, with the ML configuration showing slightly degraded spatial pattern accuracy. Column-integrated cloud water (ice + liquid) (Figure~\ref{fig:maps_tas_pr}, bottom row) from observations shows maxima in the mid-latitude storm track regions. Both schemes exhibit systematic negative biases across most oceanic regions. Both schemes show biases from $-0.06$ to $+0.06$ kg m$^{-2}$ and a low spatial pattern correlation. The ML configuration shows a slightly higher correlation and lower average RMSE value. 

Total cloud cover (Figure~\ref{fig:maps_lwp_iwp}, top row) shows observed maxima in the storm track regions and tropical convergence zones. The classical graupel scheme exhibits relatively small mean bias and moderate correlation ($+0.92\,\%$, RMSE = $8.45\,\%$, $R^2 = 0.68$). The ML configuration shows a larger mean bias ($+4.97\,\%$) with substantially higher average RMSE ($12.61\,\%$). The negative spatial correlation ($R^2 = -0.57$) of the ML configuration indicates that the observed spatial pattern of total cloud cover is not reproduced well, performing considerably worse than the classical graupel scheme in this regard. This degraded performance in total cloud cover is the largest difference between the two schemes across all evaluated variables, and is spatially consistent with the stronger positive bias in the tropical upper troposphere already identified in the zonal-mean cloud cover diagnostics in Section \ref{sec:zonal_means}. For cloud ice water path (Figure~\ref{fig:maps_lwp_iwp}, middle row), both schemes produce systematic negative biases across nearly all regions (baseline: $-0.0286$\,kg\,m$^{-2}$, RMSE = $0.0337$\,kg\,m$^{-2}$, $R^2 = -24.17$; ML: $-0.0294$\,kg\,m$^{-2}$, RMSE = $0.0347$\,kg\,m$^{-2}$, $R^2 = -31.95$). Both schemes underestimate cloud ice water path, particularly in tropical regions with frequent and strong convection and in mid-latitude storm track regions, with the ML configuration showing a marginally worse spatial correlation. We would like to emphasize, that satellite observations of cloud ice and liquid water path are subject to large uncertainties. For this reason, multiple datasets have been averaged to "MultiObsMean" means for comparison with the ICON results. Quantitative comparisons, however, remain challenging because of the large uncertainties. Cloud liquid water path (Figure~\ref{fig:maps_lwp_iwp}, bottom row) shows an observed global mean of $6.66\cdot10^{-2}$\,kg\,m$^{-2}$ (averaged across CLARA-AVHRR, CloudSat, ESACCI-CLOUD, MERRA2, and MODIS; see Table~\ref{tab:global_means_refs}), with maxima in the subtropical stratocumulus regions and mid-latitude storm track regions. Both schemes exhibit negative biases over stratocumulus regions (e.g. the subtropical eastern Pacific and Atlantic) and positive biases in some tropical areas (e.g. the tropical Indian Ocean and western Pacific), indicating well-known deficiencies in representing marine subtropical boundary layer clouds. The ML configuration introduces larger regional errors despite its smaller global mean bias ($+0.0019$\,kg\,m$^{-2}$ vs. $-0.0070$\,kg\,m$^{-2}$ for the baseline), with a higher RMSE ($0.0520$ vs. $0.0429$\,kg\,m$^{-2}$) and lower spatial correlation ($R^2 = 0.18$ vs. $R^2 = 0.35$).

The comparisons of the geographical patterns from the ICON simulations with observations reveal that the ML microphysics scheme does not improve upon the classical baseline for most variables, and specifically for cloud cover, shows a lower spatial pattern correlation. The fact that both schemes show similar bias structures for near-surface temperature and precipitation but diverge for cloud-related fields suggests that the ML configuration's learned cloud microphysics interact differently than the classical graupel scheme with the diagnostic cloud cover scheme and the radiation parameterization. We would like to emphasize, however, that in contrast to the classical graupel scheme, the ML-based scheme does not have specific tuning parameters reducing the risk of error compensation by tuning.
\section{Discussion}
% opening summary
In this work, we demonstrate stable decade-long online runs of an ML-based cloud microphysics parameterization trained on data from convection-permitting simulations coupled with the atmosphere component of the ICON model. Previous work \citep{Sarauer2024,Sarauer2025} that evaluated the ML microphysics scheme offline showed good results by being able to implicitly take into account highly resolved dynamics for the cloud microphysics. The coupled system maintains numerical stability across all seasons despite training exclusively on January data. It achieves comparable skill in simulating climate as the classical graupel scheme and shows physically consistent spatial patterns. While mean-state biases are not yet reduced relative to the classical scheme, the achievement of decade-long stability itself represents a necessary and novel milestone for this class of ML parameterizations. 

% online performance and biases
While the ML microphysics scheme maintains numerical stability over decade-long integrations, it shows some biases relative to both the classical baseline and observations. The ML configuration produces higher total cloud cover and shows somewhat degraded spatial pattern correlation for cloud-related fields, in particular total cloud cover, cloud ice water path, and cloud liquid water path. This positive bias is consistent with the corresponding negative bias in the simulated TOA shortwave cloud radiative effect. Temperature fields are nearly identical to the baseline and close to observations, while precipitation shows a small negative bias in both schemes. Both schemes systematically underestimate cloud ice water path, indicating deficiencies that extend beyond the microphysics scheme alone. The fact that biases remain relatively constant over the 10-year simulation without any drift suggests that these are rather fundamental limitations of the current ML microphysics scheme rather than numerical instabilities. Several factors likely contribute to these biases, though disentangling their individual roles is difficult. The cell-based architecture, though inspired by the architecture of the classical graupel scheme, cannot explicitly learn column-integrated conservation properties, which may contribute to the positive cloud cover bias. The results also suggest that the ML microphysics scheme interacts differently with the cloud cover diagnostic and the convection and radiation schemes than the classical parameterization. In our approach, unresolved subgrid-scale dynamics and microphysics are learned simultaneously, and it might be an easier approach to learn a unified parameterization of turbulence, convection and microphysics \citep{Lamb2026}. Also, disentangling these effects from tuning influences is non-trivial, and isolating their individual contributions would require targeted experiments beyond the scope of this study. Therefore, these points remain working hypotheses rather than demonstrated causes.

% training data and limitations
The ML model is trained on 12 days of January simulations at a 5-km resolution. This limited temporal sampling reflects the high computational cost of running global high-resolution simulations. The fact that January-only training produces stable simulations across all seasons is encouraging and suggests that the combination of the cell-based approach and physical constraints provides robustness to seasonal and inter-annual variability. Additionally, January-only training data does not capture the full range of atmospheric conditions throughout the year. Seasonal differences in temperature profiles, moisture distributions, and convective regimes are not fully represented, which likely contributes to the biases discussed above. This is consistent with the higher tuning loss observed in July. Expanding training data to include multiple seasons and possibly multiple years is the next step, and given the strong relationship between training data volume and model performance demonstrated in Section~\ref{subssec:stability}, this could reduce systematic biases.

% computational efficiency
The current implementation increases the total simulation time needed by ICON by approximately 3$\times$ compared to the classical graupel scheme (one simulation year requires 4.5 hours versus 1.5 hours on 12 nodes on the HPC system at DKRZ - 2 x AMD 7763 CPU per node with 64 cores each). This reflects a 155$\times$ slowdown in the microphysics component itself (see more details in \ref{appendix_time}). Since microphysics represents a small fraction of total runtime, this translates to a 3$\times$ increase in total. The overhead stems from cell-based processing and unoptimized inference through the \texttt{ftorch} interface \citep{Atkinson_2025}. Promising optimization pathways include column-based batched inference and model compression. For context, \cite{Perkins_Brenowitz_Bretherton_Nugent_2024} achieved 18$\times$ speedup when emulating microphysics, demonstrating that ML configurations can be competitive when performance is prioritized. Given the challenges of achieving stable online coupling, prioritizing stability over speed was appropriate for this initial demonstration. However, reducing computational overhead is essential for practical applications in climate projections.

% tunability
The current ML microphysics scheme is trained without explicit tunable parameters, limiting the ability to adjust model behavior post-training in the scheme itself as it is routinely done with classical graupel schemes. As tuning parameters are typically only weakly constrained, they can lead to unintentional error compensation, where a model produces apparently good results for the wrong physical reasons. The absence of such parameters in the ML microphysics scheme can therefore be seen as an advantage, as the risk of this kind of error compensation is expected to be reduced. Recent work has explored approaches where ML components include interpretable, tunable parameters that can be adjusted within the coupled model \citep{Balogh_2022} or where the ML microphysics scheme is mixed with the conventional one when the ML model is uncertain \citep{Heuer2025}. This combines the power of neural networks with the flexibility of traditional parameterizations. Adding this kind of tunability to future ML microphysics schemes could make it easier to calibrate models and reduce the gap between the applicability of ML-based and classical graupel schemes.

%Generalization to future climates
Our evaluation of the ICON results with observations demonstrates spatial and seasonal generalization for the present climate, but we have not tested behavior under climate change scenarios. An important question is how the parameterization performs in significantly warmer or cooler climates, such as $\pm$4K scenarios commonly used to assess a model's response to greenhouse gas forcing. Classical parameterizations rely on physical principles that should hold across temperature ranges, though retuning might be necessary. On the other hand, ML microphysics schemes trained on present-day climate data may show unpredictable behavior when temperature and moisture distributions shift significantly. This would be the case if atmospheric states fall outside the training distribution. Because our model was trained on high-resolution convection-permitting data, we expect it to handle a broader range of atmospheric conditions reasonably well. However, only systematic testing in perturbed climate experiments can confirm whether a model is suitable for long-term climate projections and whether training data for warmer or cooler conditions would be needed.

%Aerosol-cloud interactions
The current implementation uses fixed cloud droplet number concentrations and does not account for aerosol-cloud interactions, which remain a large source of uncertainty in climate projections \citep{IPCC_2023}. A natural next step is to develop an aerosol-aware ML microphysics scheme by incorporating aerosol effects, e.g. by introducing cloud droplet number concentration (CDNC) and possibly ice crystal number concentration as additional input features to the ML microphysics scheme. While prognostic aerosols in a high-resolution model are still computationally expensive, recent progress suggests this is becoming more feasible \citep{Klocke_2025}, and such aerosol-aware training simulations would include implicit signatures of aerosol effects that an ML microphysics scheme could learn and transfer into coarse-resolution models.

% lessons learned
We conclude that stable online coupling cannot be inferred from offline performance alone, additional constraints are needed. Mass positivity emerged as the most critical constraint, while overshoot prevention was essential for decade-long stability. Physical constraints that appear redundant during training become indispensable safeguards in the coupled system, where rare tail events and distributional shifts can trigger numerical instabilities. The iterative testing turned out to be a necessity: instabilities invisible in offline evaluation only manifested after days or weeks of integration. Most existing ML microphysics work targets emulation at the same resolution \citep{gettelman_2021, Perkins_Brenowitz_Bretherton_Nugent_2024}, aiming for computational acceleration rather than improved physical representation. Our parameterization instead transfers information from convection-permitting simulations at 5\,km to the coarse-resolution model at 80\,km, and to our knowledge represents the first demonstration of stable, globally coupled ML cloud microphysics including ice processes, derived from convection-permitting training data over a decade-long integration. Since only long, stable coupled runs make such an assessment possible at all, our work also shows that this approach needs further development to improve upon the conventional parameterization, for example by developing a unified parameterization of turbulence, convection and microphysics \citep{Lamb2026}, or by integrating tuning parameters into the ML microphysics scheme.

More broadly, this work challenges the notion that ML microphysics schemes are incompatible with rigorous climate modeling. The key contribution of our work is therefore not a new ML architecture, but the demonstration that carefully designed physical constraints, data curation, and online testing are necessary requirements for stable long-term coupling, providing a foundation from which systematic improvements can be pursued. Physics understanding is thus needed to make ML methods work for climate modeling, and the stable, fully evaluated baseline presented here provides the community with a concrete starting point for those improvements.

\appendix
\section{Result Figures}\label{result_figures}
\begin{figure*}[htb]
    \centering
    \includegraphics[width=\textwidth]{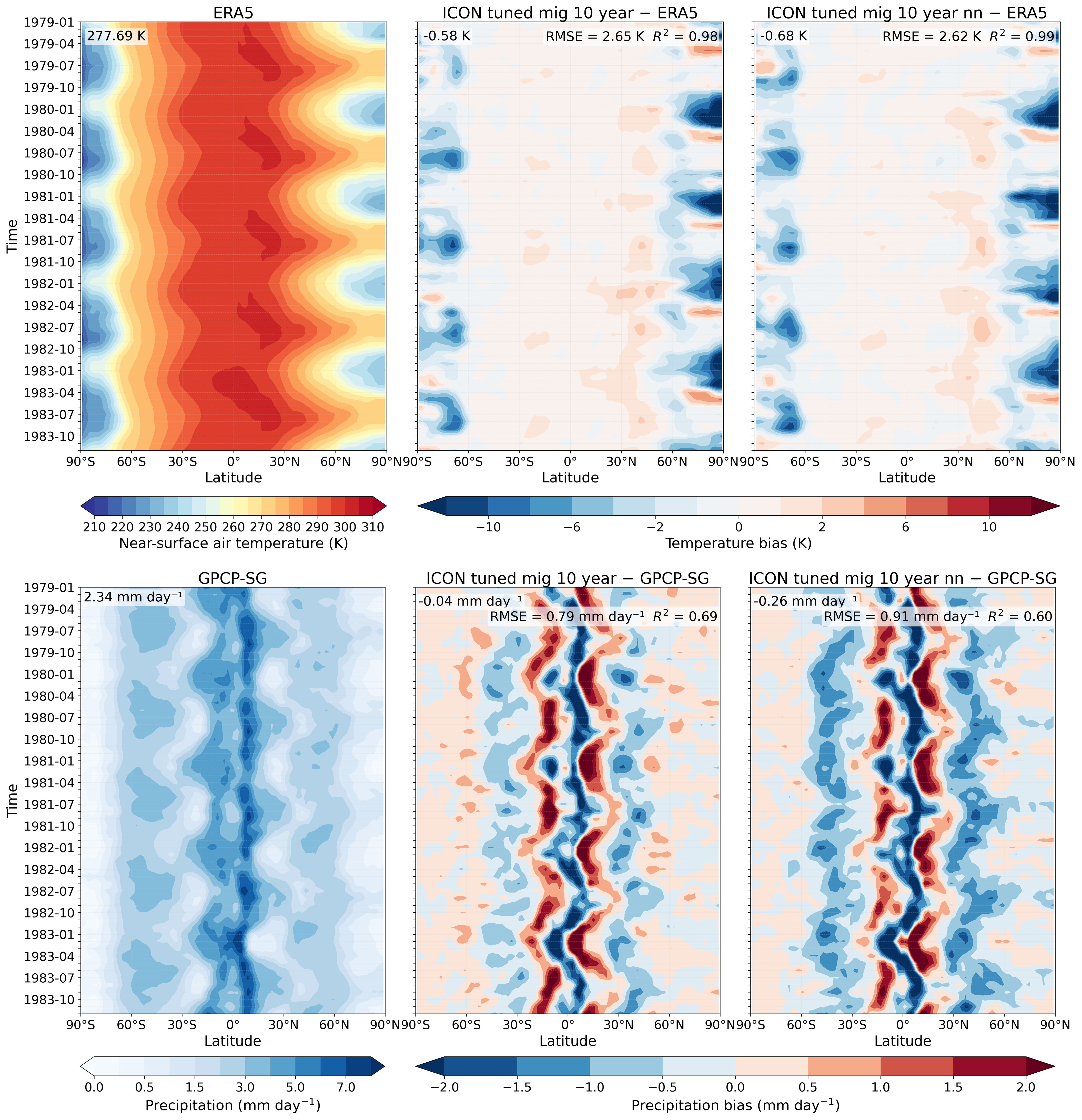}
    \caption{Global time–latitude (Hovmöller) diagrams of zonal-mean precipitation (pr) and near-surface air temperature (tas). The left column shows the reference observational or reanalysis products (ERA5 for tas and GPCP-SG for pr). The middle and right columns show differences between ICON simulations and the corresponding reference data for the tuned configuration without (ICON\_tuned\_mig\_10\_year) and with (ICON\_tuned\_mig\_10\_year\_nn) the ML microphysics scheme, respectively. Numbers on the top of each bias panel indicate the global mean bias, RMSE, and spatial correlation ($R^2$) with respect to the reference dataset.}
    \label{fig:hovmoller}
\end{figure*}

\begin{figure*}[htb]
    \centering
    \includegraphics[width=\textwidth]{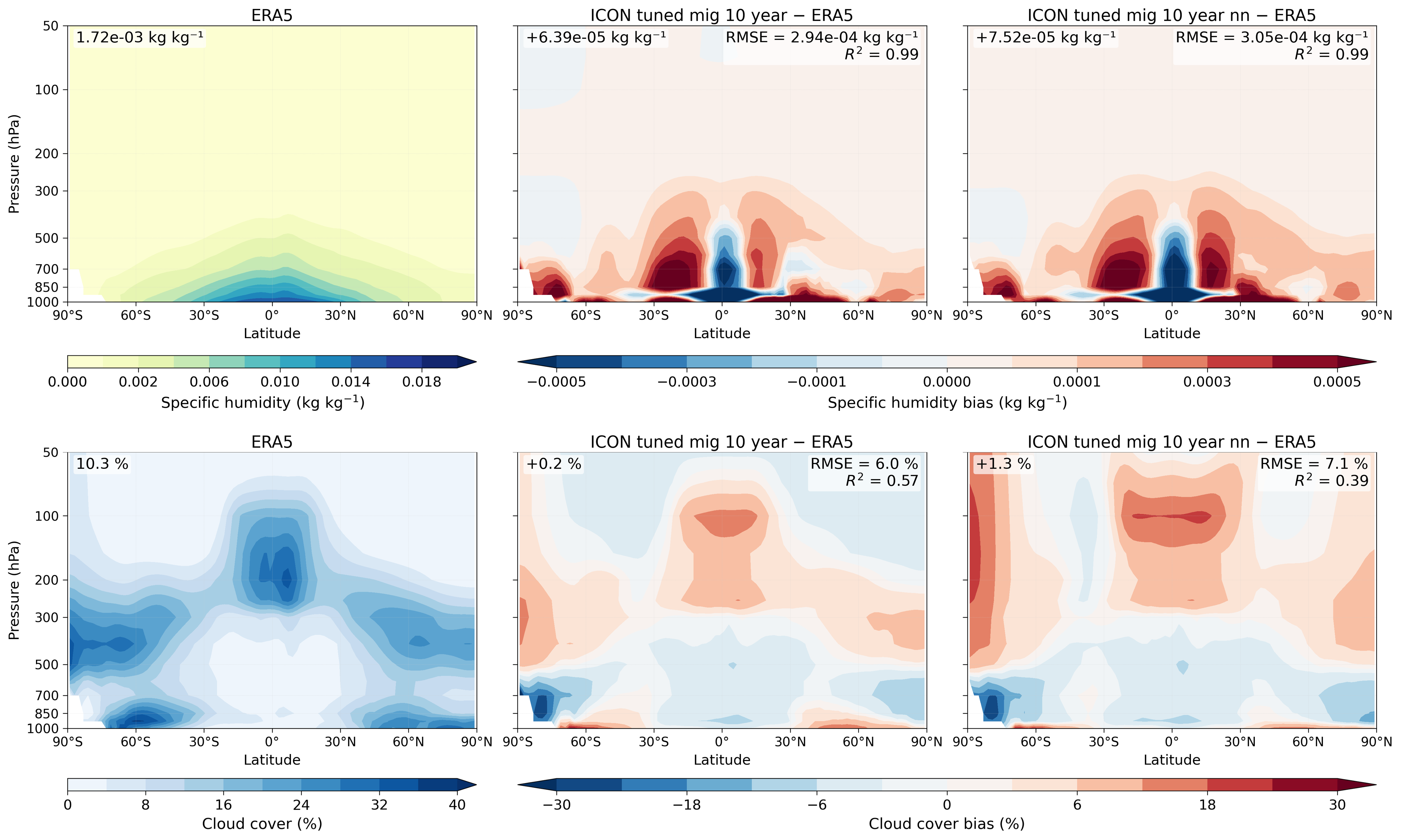}
    \caption{
    Zonal annual mean vertical profiles averaged over the 10-year simulation period. The top row shows specific humidity, the bottom row cloud fraction.
    Observational reference data are shown in the left column (ERA5).
    The center column shows differences between ERA5 and the tuned ICON baseline simulation using the classical graupel microphysics scheme, the right column shows differences between ERA5 and the tuned ICON simulation using the ML-based cloud microphysics parameterization. Numbers on the top of each bias panel indicate the global mean bias, RMSE, and spatial correlation ($R^2$) with respect to the reference dataset.
    }
    \label{fig:vertical_humidity_cloudcover}
\end{figure*}

\begin{figure*}[htb]
    \centering
    \includegraphics[width=\textwidth]{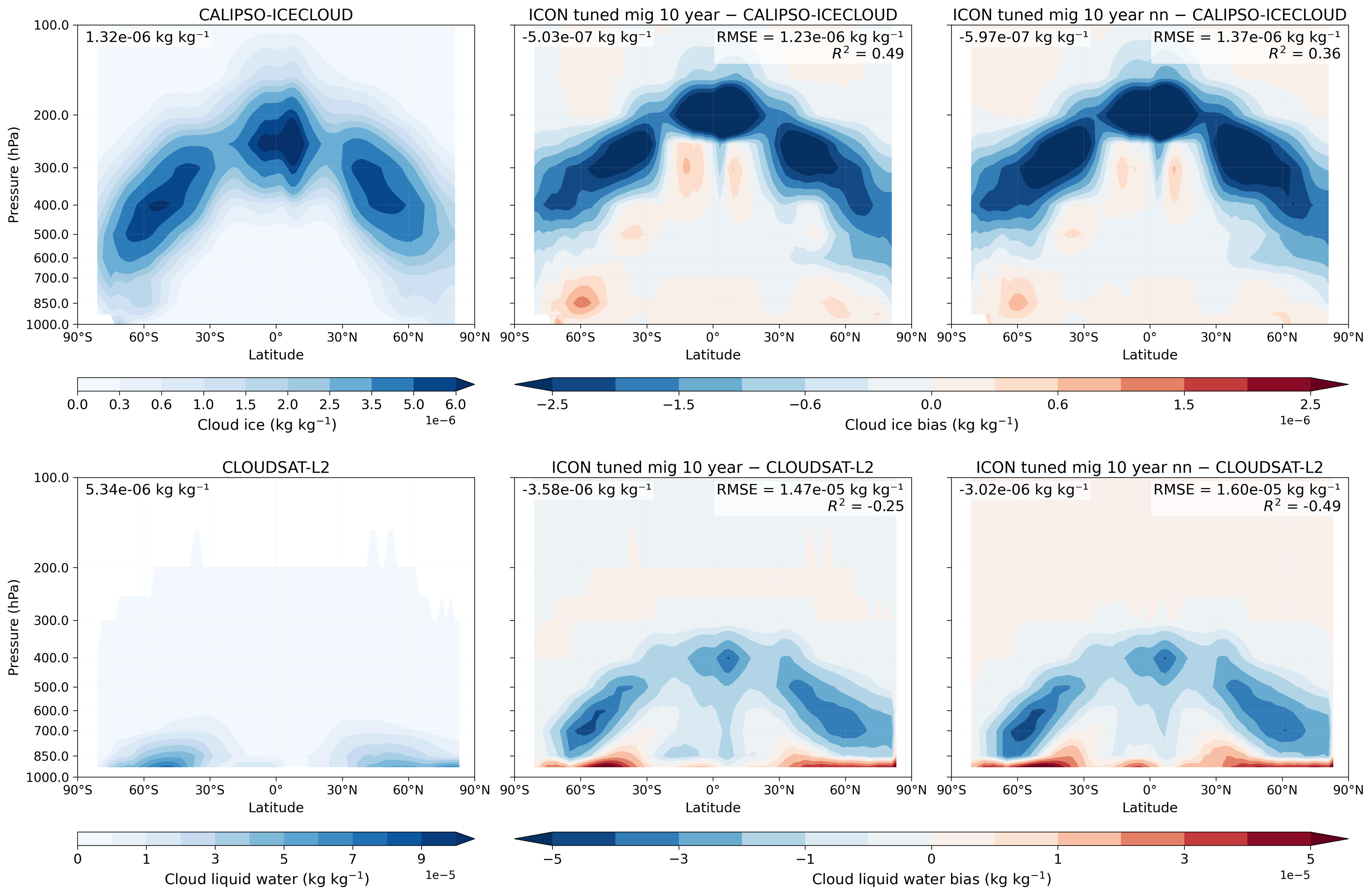}
    \caption{
    Zonal-mean vertical structure of the cloud water (liquid and ice) content averaged over the 10-year simulation period.
    The top row shows cloud ice, and the bottom row shows cloud liquid water as a function of latitude and pressure.
    Observational reference data from CALIPSO-ICECLOUD \citep{NASA/LARC/SD/ASDCCALIPSO} for cloud ice and from CloudSat \citep{Stephens_etal_2002, Stephens_etal_2018}for cloud liquid water are shown in the left column.
    The center column shows differences between the tuned ICON baseline simulation using the classical graupel microphysics scheme and CALIPSO-ICECLOUD, the right column shows differences between the tuned ICON simulation using the ML-based cloud microphysics parameterization and CloudSat data. Numbers on the top of each bias panel indicate the global mean bias, RMSE, and spatial correlation ($R^2$) with respect to the reference dataset.   
    }
    \label{fig:vertical_cloud_condensate}
\end{figure*}

\begin{figure*}[h]
    \centering
    \includegraphics[width=\textwidth]{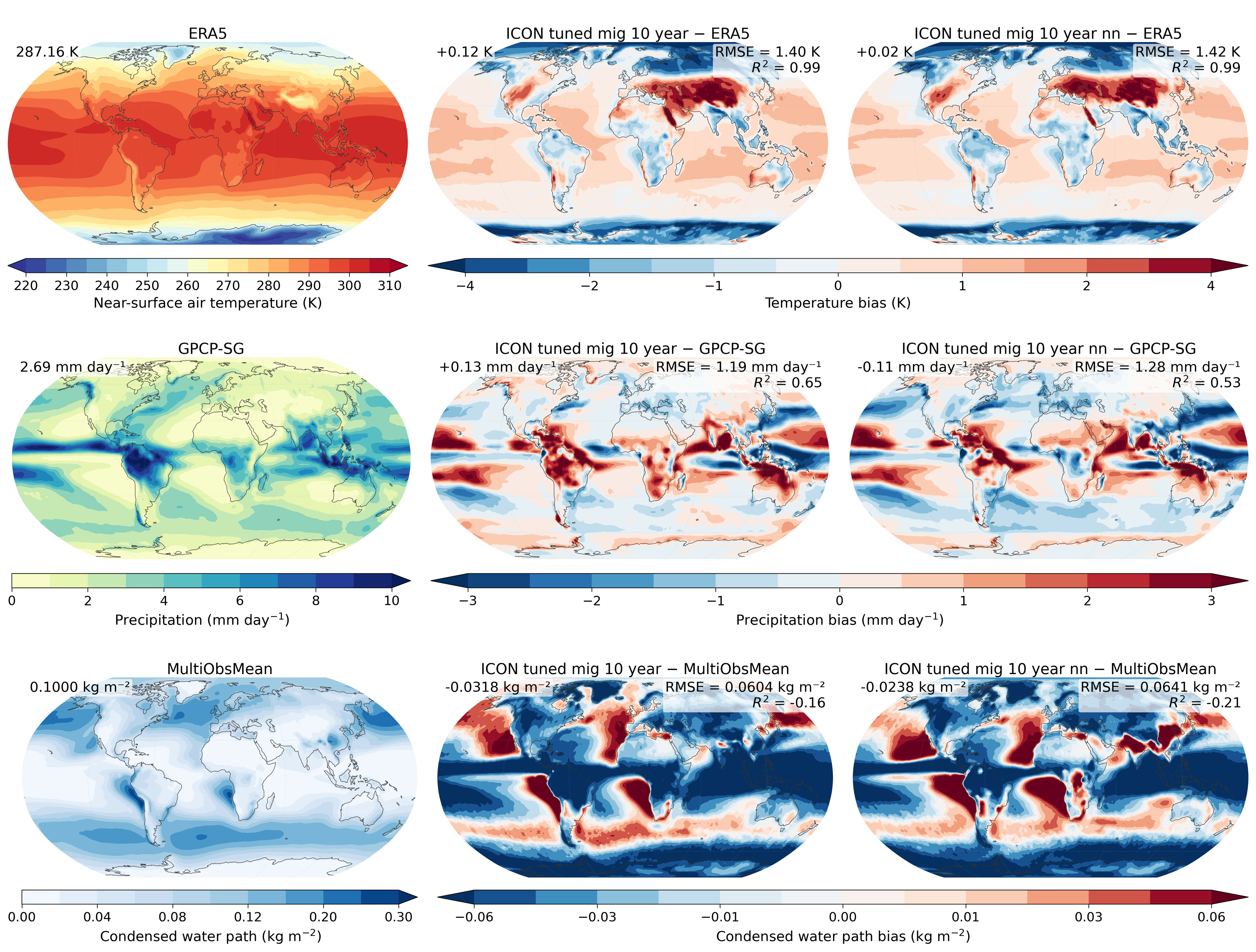}
    \caption{Maps of 10-year annual means and model biases for near-surface air temperature (tas), precipitation (pr), and condensed water path (clwvi). The left column shows reference observational or reanalysis products (ERA5  for tas, GPCP-SG for pr, and MultiObsMean (CLARA-AVHRR, CloudSat, ESACCI-CLOUD, MERRA2, MODIS; see Table~\ref{tab:global_means_refs}) for clwvi). The middle and right columns show differences between ICON simulations and the corresponding reference data for the tuned configuration with (ICON\_tuned\_mig\_10\_year\_nn) and without (ICON\_tuned\_mig\_10\_year) ML microphysics scheme. Numbers on the top of each bias panel indicate the global mean bias, RMSE, and spatial correlation ($R^2$) with respect to the reference dataset.}
    \label{fig:maps_tas_pr}
\end{figure*}

\begin{figure*}[h]
    \centering
    \includegraphics[width=\textwidth]{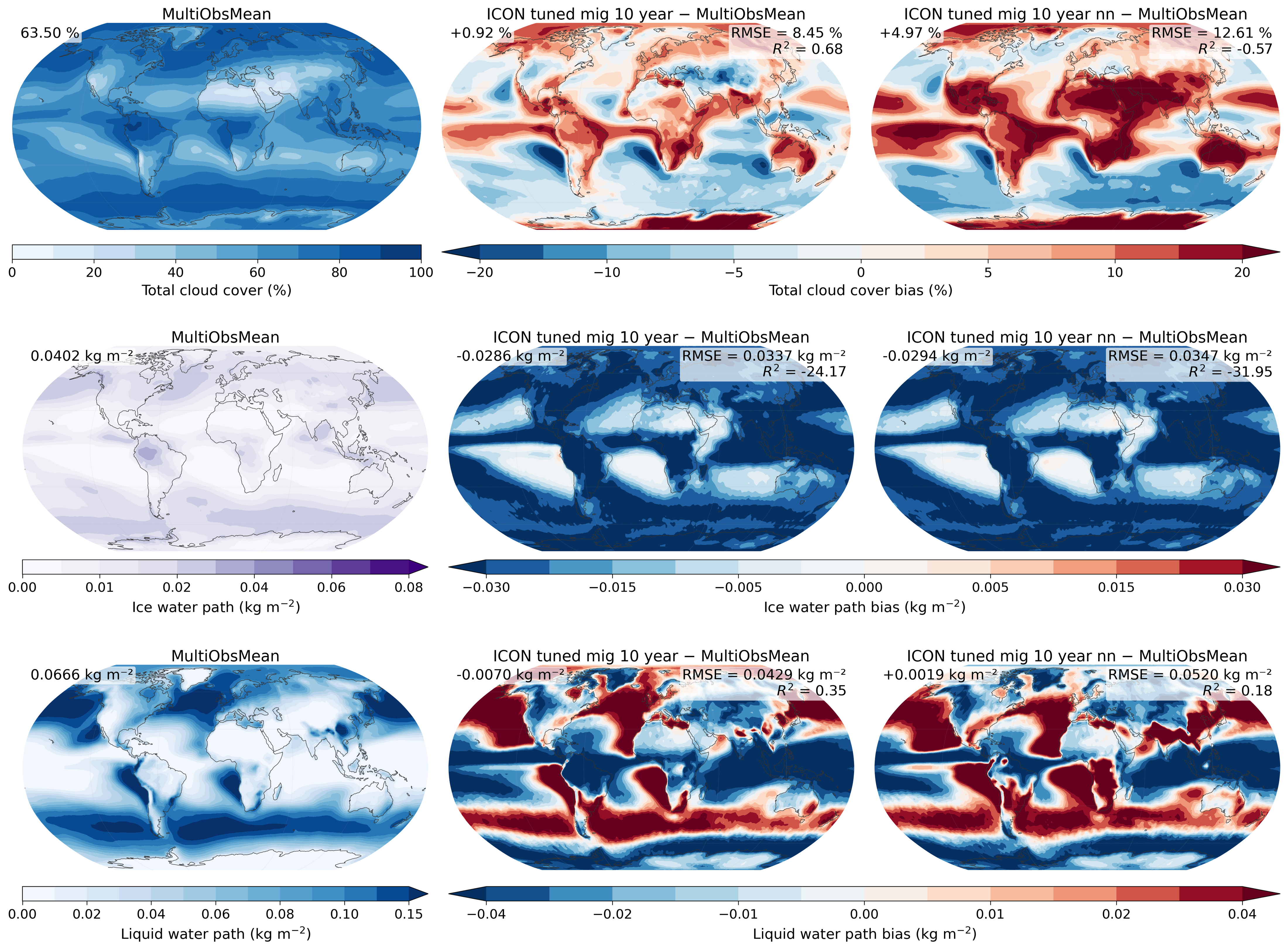}
    \caption{Global annual mean climatologies and model biases for cloud cover (clt), ice water path (clivi), and liquid water path (lwp). The left column shows reference observational products as MultiObsMean (clt: CLARA-AVHRR, ESACCI-CLOUD, MERRA2, MODIS, PATMOS-x; clivi: CLARA-AVHRR, CloudSat, ESACCI-CLOUD, MERRA2, MODIS; lwp: CLARA-AVHRR, CloudSat, ESACCI-CLOUD, MERRA2, MODIS; see Table~\ref{tab:global_means_refs}). The middle and right columns show differences between ICON simulations and the corresponding reference data for the tuned configuration with (ICON\_tuned\_mig\_10\_year\_nn) and without (ICON\_tuned\_mig\_10\_year) ML microphysics scheme. Numbers in the top left of each bias panel indicate the global mean bias, RMSE, and spatial correlation ($R^2$) with respect to the reference dataset.}
    \label{fig:maps_lwp_iwp}
\end{figure*}

The following figures correspond to Section \ref{sec:evaluation}. Figure~\ref{fig:hovmoller} shows Hovmöller diagrams of zonal-mean precipitation and near-surface air temperature for both ICON configurations and the corresponding observational reference data. Figure~\ref{fig:vertical_humidity_cloudcover} shows zonal-mean vertical profiles of specific humidity and cloud fraction, and Figure~\ref{fig:vertical_cloud_condensate} shows the corresponding profiles of cloud ice and cloud liquid water content. Figures~\ref{fig:maps_tas_pr} and~\ref{fig:maps_lwp_iwp} show global maps of 10-year annual mean biases for near-surface temperature, precipitation, condensed water path, cloud cover, ice water path, and liquid water path.

\section{Tendency Distributions and Outlier Sampling}\label{appendix_distribution}

\begin{figure}[htb]
    \centering
    \includegraphics[width=\textwidth]{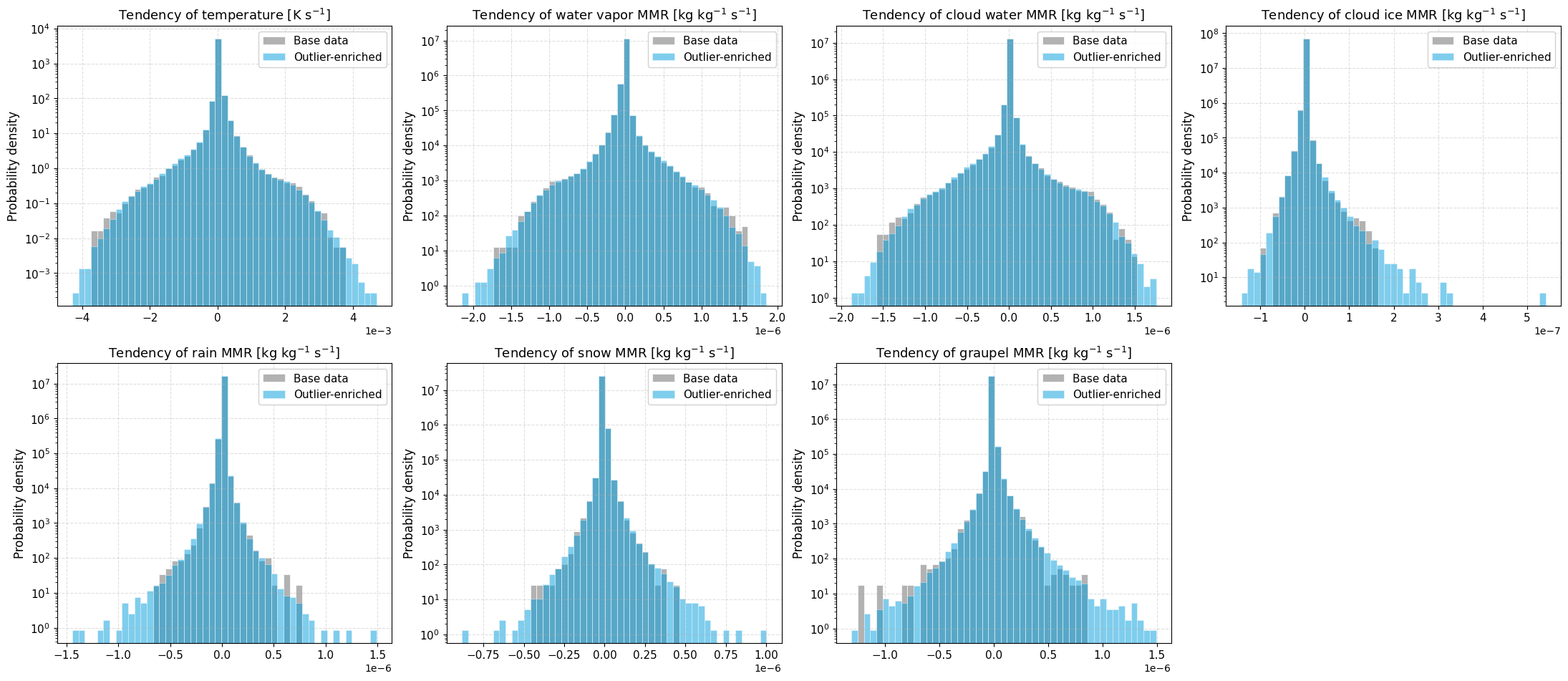} \vspace{3mm}
    \caption{\label{fig:outlier} Distributions of microphysical tendencies for all seven output variables on a logarithmic count axis. Gray bars show the base dataset, blue bars the outlier-enriched dataset.}
\end{figure}

Figure~\ref{fig:outlier} shows the distributions of all seven microphysical 
tendency outputs for the base and outlier-enriched datasets. All variables show a distinct peak near zero, with the base dataset containing very few samples in the distribution tails. The outlier-enriched dataset increases the representation of extreme tendency values through the quantile-based approach described in Section~\ref{sec:outlier}, while preserving the overall shape of the distribution.

\section{Stability}\label{appendix_stability}
\begin{figure*}[htb]
    \centering
    \includegraphics[width=\textwidth]{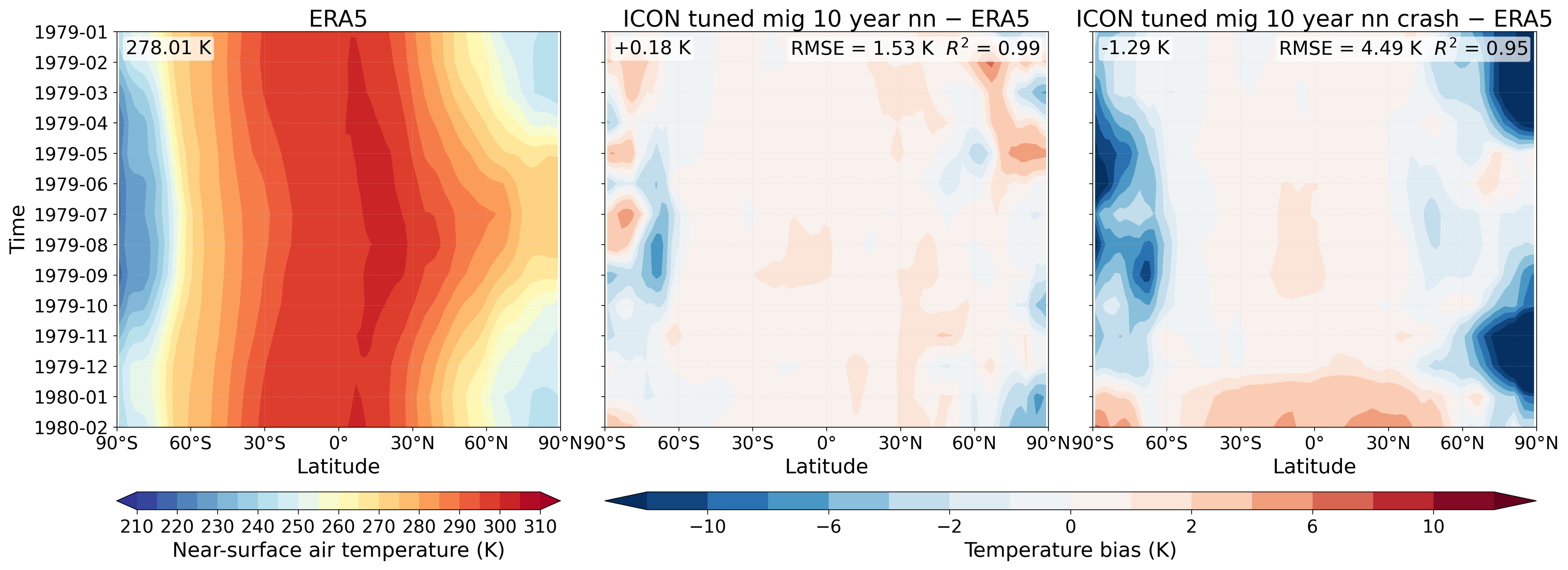}
    \caption{Global time–latitude (Hovmöller) diagrams of zonal-mean near-surface air temperature (tas). The left column shows ERA5 reanalysis as reference. The middle column shows the bias of the stable ML configuration (ICON\_tuned\_mig\_10\_year\_nn) relative to ERA5, and the right column shows the corresponding bias for an unstable ML configuration (ICON\_tuned\_mig\_10\_year\_nn\_crash). Numbers on the top of each bias panel indicate the global mean bias, RMSE, and spatial correlation ($R^2$) with respect to the reference dataset.}
    \label{fig:unstable_hovmoller}
\end{figure*}

Figure~\ref{fig:unstable_hovmoller} illustrates an early ML configuration that became unstable and crashed after approximately 1.3 years of simulation (5 Million sample configuration in Table \ref{tab:stability}). The instability can be seen by the gradual development of temperature biases: cold biases emerge and intensify at high latitudes in both hemispheres, while warm biases increase in the Tropics. These opposite bias patterns amplify over time and eventually lead to numerical instability and model failure.

This configuration was trained on insufficient data and did not generalize robustly beyond the training conditions. After increasing the training data volume, all subsequent configurations remained stable over multi-year and decade-long simulations, and no further unstable runs were observed. The stable configuration used throughout the manuscript does not exhibit such behavior.
\section{Evaluation variable names}\label{appendix_names}
Table~\ref{tab:cmip6_vars} provides a reference list of all variables evaluated with ESMValTool, including their "short names" and descriptions.
\begin{table}[htb]
\centering
\caption{CMIP6 variable names and their descriptions as used in this study}
\label{tab:cmip6_vars}
\begin{tabular}{lll}
\hline
\textbf{Short name} & \textbf{Description} & \textbf{Unit} \\
\hline
asr    & Absorbed shortwave radiation at TOA          & W\,m$^{-2}$ \\
clivi  & Cloud ice water path (column-integrated)     & kg\,m$^{-2}$ \\
clt    & Total cloud cover                            & \% \\
clwvi  & Condensed water path (cloud water + ice)     & kg\,m$^{-2}$ \\
hus400 & Specific humidity at 400\,hPa                & kg\,kg$^{-1}$ \\
lwcre  & Longwave cloud radiative effect at TOA       & W\,m$^{-2}$ \\
lwp    & Cloud liquid water path                      & kg\,m$^{-2}$ \\
pr     & Precipitation                                & mm\,day$^{-1}$ \\
prw    & Water vapor path (column-integrated)         & kg\,m$^{-2}$ \\
rlut   & Outgoing longwave radiation at TOA           & W\,m$^{-2}$ \\
rsut   & Outgoing shortwave radiation at TOA          & W\,m$^{-2}$ \\
swcre  & Shortwave cloud radiative effect at TOA      & W\,m$^{-2}$ \\
ta200  & Air temperature at 200\,hPa                 & K \\
ta850  & Air temperature at 850\,hPa                 & K \\
tas    & Near-surface air temperature                 & K \\
tauu   & Eastward wind stress                         & N\,m$^{-2}$ \\
ua200  & Eastward wind at 200\,hPa                   & m\,s$^{-1}$ \\
ua850  & Eastward wind at 850\,hPa                   & m\,s$^{-1}$ \\
\hline
\end{tabular}
\end{table}

\section{Observational reference datasets for model evaluation}
\label{appendix_obs_datasets}
Table~\ref{tab:global_means_refs} lists the observational and reanalysis 
datasets used to compute the observed mean values shown in 
Table~\ref{tab:global_means} and the reference data shown in the zonal 
mean profile plots. The observed mean in Table~\ref{tab:global_means} is 
calculated as the average across all listed datasets for each variable.

\begin{table}[htb]
\centering
\small
\caption{Observational reference datasets used for the global mean climate 
variables (Table~\ref{tab:global_means}), Hovmöller diagrams and global maps 
(Figures~\ref{fig:hovmoller},~\ref{fig:maps_tas_pr} and~\ref{fig:maps_lwp_iwp}), 
and zonal mean profile plots (Figures~\ref{fig:vertical_humidity_cloudcover} 
and~\ref{fig:vertical_cloud_condensate}).}
\label{tab:global_means_refs}
\begin{tabular}{p{0.45\textwidth}p{0.55\textwidth}}
\hline
\textbf{Variable} & \textbf{Reference dataset(s)} \\
\hline
\multicolumn{2}{l}{\it Global mean (Table~\ref{tab:global_means}), Hovmöller diagrams and global maps (Figures~\ref{fig:hovmoller},~\ref{fig:maps_tas_pr} and~\ref{fig:maps_lwp_iwp})} \\
\hline
Near-surface air temperature (K) & ERA5 \citep{Hersbach_2020} \\
Precipitation (mm\,day$^{-1}$) & GPCP-SG \citep{Adler_etal_2018} \\
Water vapor path (kg\,m$^{-2}$) & ERA5 \citep{Hersbach_2020}, ESACCI-WATERVAPOUR \citep{Schroeder_etal_2023}, ISCCP-FH \citep{Rossow_etal_2016, Young_etal_2018}, MERRA2 \citep{Gelaro_etal_2017} \\
Total cloud cover (\%) & CLARA-AVHRR \citep{Karlsson_etal_2017, Karlsson_etal_2020}, ERA5 \citep{Hersbach_2020}, ESACCI-CLOUD \citep{Stengel_etal_2020}, MERRA2 \citep{Gelaro_etal_2017}, MODIS \citep{Platnick_etal_2003}, PATMOS-x \citep{Heidinger_etal_2014} \\
Cloud liquid water path ($10^{-2}\,$kg m$^{-2}$) & CLARA-AVHRR \citep{Karlsson_etal_2017, Karlsson_etal_2020}, CloudSat \citep{Stephens_etal_2002, Stephens_etal_2018}, ERA5 \citep{Hersbach_2020}, ESACCI-CLOUD \citep{Stengel_etal_2020}, MERRA2 \citep{Gelaro_etal_2017}, MODIS \citep{Platnick_etal_2003} \\
Cloud ice water path ($10^{-2}\,$kg\,m$^{-2}$) & CLARA-AVHRR \citep{Karlsson_etal_2017, Karlsson_etal_2020}, CloudSat \citep{Stephens_etal_2002, Stephens_etal_2018}, ERA5 \citep{Hersbach_2020}, ESACCI-CLOUD \citep{Stengel_etal_2020}, MERRA2 \citep{Gelaro_etal_2017}, MODIS \citep{Platnick_etal_2003} \\
Shortwave cloud radiative effect (W\,m$^{-2}$) & CERES-EBAF \citep{NASA/LARC/SD/ASDC2022CERES, Loeb_etal_2009, Loeb_etal_2012}, ESACCI-CLOUD \citep{Stengel_etal_2020}, ISCCP-FH \citep{Rossow_etal_2016, Young_etal_2018}, MERRA2 \citep{Gelaro_etal_2017} \\
Longwave cloud radiative effect (W\,m$^{-2}$) & CERES-EBAF \citep{NASA/LARC/SD/ASDC2022CERES, Loeb_etal_2009, Loeb_etal_2012}, ERA5 \citep{Hersbach_2020}, ESACCI-CLOUD \citep{Stengel_etal_2020}, ISCCP-FH \citep{Rossow_etal_2016, Young_etal_2018}, MERRA2 \citep{Gelaro_etal_2017} \\
\hline
\multicolumn{2}{l}{\it Zonal mean profiles (Figures~\ref{fig:vertical_humidity_cloudcover} and~\ref{fig:vertical_cloud_condensate})} \\
\hline
Specific humidity (kg\,kg$^{-1}$) & ERA5 \citep{Hersbach_2020} \\
Cloud cover (\%) & ERA5 \citep{Hersbach_2020} \\
Cloud ice mass mixing ratio (kg\,kg$^{-1}$) & CALIPSO-ICECLOUD \citep{NASA/LARC/SD/ASDCCALIPSO} \\
Cloud liquid water mass mixing ratio (kg\,kg$^{-1}$) & CloudSat \citep{Stephens_etal_2002, Stephens_etal_2018} \\
\hline
\end{tabular}
\end{table}

\section{Computational Cost}\label{appendix_time}
We assess computational cost using ICON's built-in timing diagnostics on 12-node simulations. The ML microphysics component itself is 155$\times$ slower than the classical graupel scheme (Table~\ref{tab:timing}), but, since microphysics represents only a small fraction of the total computational time for a simulation, the overall slowdown is approximately 3$\times$: one simulation year requires 1.5 hours with the classical graupel scheme versus 4.5 hours with the ML configuration.

\begin{table}[htb]
\centering
\caption{Computational cost of cloud microphysics schemes. Times shown are total wall-clock time for the microphysics component over a 10-year simulation on 12 nodes.}
\label{tab:timing}
\begin{tabular}{lrr}
\hline
\textbf{Scheme} & \textbf{Microphysics time (s)} & \textbf{Relative cost} \\
\hline
Classical Graupel & 76.1 & 1.0$\times$ \\
ML Microphysics & 11,806 & 155$\times$ \\
\hline
\end{tabular}
\end{table}

The large slowdown of the ML-based cloud microphysics arises from several factors. First, the classical graupel scheme is computationally simple, consisting primarily of algebraic expressions and lookup tables. Second, the current ML implementation uses the \texttt{ftorch} interface \citep{Atkinson_2025} without HPC optimization. Third, the model processes grid cells individually rather than using batched inference. Fourth, physical constraint layers add per-cell overhead. This cost is acceptable for work focused on stability, but operational use would require optimization. Several pathways exist for potentially substantial speedups: 
\begin{itemize}
    \item model compression through quantization or pruning,
    \item batched inference across vertical columns,
    \item GPU acceleration.
\end{itemize}
\cite{Perkins_Brenowitz_Bretherton_Nugent_2024} achieved an 18$\times$ speedup when emulating a more complex microphysics scheme, demonstrating that ML microphysics schemes can be competitive when computational optimization is prioritized. For this study, we prioritized long-term stability and physical consistency over computational efficiency. Future work optimizing inference would be necessary before operational application of this scheme in climate models.

%%%%%%%%%%%%%%%%%%%%%%%%%%%%%%%%%%%%%%%%%%%%%%%
% Optional Glossary, Notation or Acronym section goes here:
%
% Glossary is only allowed in Reviews of Geophysics
%  \begin{glossary}
%  \term{Term}
%   Term Definition here
%  \term{Term}
%   Term Definition here
%  \term{Term}
%   Term Definition here
%  \end{glossary}

%%%%%%%%%%%%%%%%%%%%%%%%%%%%%%%%%%%%%%%%%%%%%%%
% Acronyms
%% NOTE that acronyms in the final published version will be spelled out when used in figure captions.
%   \begin{acronyms}
%   \acro{Acronym}
%   Definition here
%   \acro{EMOS}
%   Ensemble model output statistics
%   \acro{ECMWF}
%   Centre for Medium-Range Weather Forecasts
%   \end{acronyms}

%%%%%%%%%%%%%%%%%%%%%%%%%%%%%%%%%%%%%%%%%%%%%%%
% Notation
%   \begin{notation}
%   \notation{$a+b$} Notation Definition here
%   \notation{$e=mc^2$}
%   Equation in German-born physicist Albert Einstein's theory of special
%  relativity that showed that the increased relativistic mass ($m$) of a
%  body comes from the energy of motion of the body—that is, its kinetic
%  energy ($E$)—divided by the speed of light squared ($c^2$).
%   \end{notation}

%%%%%%%%%%%%%%%%%%%%%%%%%%%%%%%%%%%%%%%%%%%%%%%
%
% DATA SECTION and ACKNOWLEDGMENTS
%
%%%%%%%%%%%%%%%%%%%%%%%%%%%%%%%%%%%%%%%%%%%%%%%

\section*{Open Research Section}
The Machine Learning code for this work will be published under \url{https://github.com/EyringMLClimateGroup/sarauer25mlearth_ml_microphysics_parameterization} and archived on Zenodo upon acceptance. The simulation data used to train and evaluate the ML algorithms was generated with the ICON model (version 2.6.7 at 5\,km resolution for training and version 2.6.4 at 80\,km resolution for the coupled simulations). The source code is available on the \mbox{GitLab} of the Deutsches Klimarechenzentrum (DKRZ, \url{https://gitlab.dkrz.de/icon/icon-mpim}) under a BSD \mbox{3-clause} license (\url{https://gitlab.dkrz.de/icon/icon-mpim/-/tree/master/LICENSES}). The simulations were performed with the branch \mbox{feature-nextgems-aerosol-microphysics} at commit 260364f1.

\acknowledgments
This project was made possible by the DLR Quantum Computing Initiative and the Federal Ministry for Research, Technology and Space; \url{qci.dlr.de/projects/klim-qml}. We acknowledge funding by the European Research Council (ERC) Synergy Grant “Understanding and Modelling the Earth System with Machine Learning (USMILE)” under the Horizon 2020 research and innovation program (Grant agreement No. 855187). V.E. was additionally supported by the Deutsche Forschungsgemeinschaft (DFG, German Research Foundation) through the Gottfried Wilhelm Leibniz Prize awarded to V.E. (Reference No. EY 22/2-1). P.S and P.W. acknowledge funding from the Horizon 2020 projects nextGEMS under grant agreement number 101003470. P.S. additionally acknowledges funding from European Union’s Horizon Europe projects CleanCloud and Embed2Scale with grant agreements 101137639 and 101131841 and their UKRI underwrite. This work used resources of the Deutsches Klimarechenzentrum (DKRZ) granted by its Scientific Steering Committee (WLA) under project IDs bd854 and bd1179. The authors also gratefully acknowledge the Earth System Modelling Project (ESM) for funding this work by providing computing time on the ESM partition of the supercomputer JUWELS \cite{JUWELS} at the Jülich Supercomputing Centre (JSC).

%%%%%%%%%%%%%%%%%%%%%%%%%%%%%%%%%%%%%%%%%%%%%%%
% REFERENCES and BIBLIOGRAPHY
%
% \bibliography{<name of your .bib file>} don't specify the file extension
% don't specify bibliographystyle
%
%%%%%%%%%%%%%%%%%%%%%%%%%%%%%%%%%%%%%%%%%%%%%%%

\bibliography{bibliography}

%Reference citation instructions and examples:
%
% Please use ONLY \cite and \citeA for reference citations.
% \cite for parenthetical references
% ...as shown in recent studies (Simpson et al., 2019)
% \citeA for in-text citations
% ...Simpson et al. (2019) have shown...
%
%
%...as shown by \citeA{jskilby}.
%...as shown by \citeA{lewin76}, \citeA{carson86}, \citeA{bartoldy02}, and \citeA{rinaldi03}.
%...has been shown \cite{jskilbye}.
%...has been shown \cite{lewin76,carson86,bartoldy02,rinaldi03}.
%... \cite <i.e.>[]{lewin76,carson86,bartoldy02,rinaldi03}.
%...has been shown by \cite <e.g.,>[and others]{lewin76}.
%
% apacite uses < > for prenotes and [ ] for postnotes
% DO NOT use other cite commands (e.g., \citet, \citep, \citeyear, \nocite, \citealp, etc.).
%

\end{document}